\documentclass[reqno,final,11pt]{amsart}
\usepackage{natbib}  
\usepackage{fancyhdr} 
\usepackage{color} 
\usepackage{hyperref} 
\usepackage{graphicx} 
\usepackage{geometry}
\usepackage[utf8]{inputenc}
\usepackage[T1]{fontenc}
\usepackage{verbatim}

\usepackage{tikz}
\usetikzlibrary{calc,shapes,backgrounds,arrows}
\usepackage{amssymb}

\definecolor{aleacolor}{rgb}{0.16,0.59,0.78}

\hypersetup{
breaklinks,
colorlinks=true,
linkcolor=aleacolor,
urlcolor=aleacolor,
citecolor=aleacolor}

\renewcommand{\cite}{\citet}

\theoremstyle{plain}

\theoremstyle{definition}

\theoremstyle{remark}

\makeatletter \@addtoreset{equation}{section} \makeatother
\renewcommand\theequation{\thesection.\arabic{equation}}
\renewcommand\thefigure{\thesection.\arabic{figure}}
\renewcommand\thetable{\thesection.\arabic{table}}

\usepackage{longtable,tabularx,booktabs}
\usepackage{subcaption,float}
\usepackage{appendix,ulem}

\begin{document}

\begin{sloppy}

\title[Tables with Critical Values for the Meta-Analysis of Genuine and Fake $\boldsymbol{p}$-Values]{Tables with Critical Values for the Meta-Analysis of Genuine and Fake $\boldsymbol{p}$-Values}

\author{Rui Santos}
\address{Escola Superior de Tecnologia e Gest\~ao, Instituto Polit\'ecnico de Leiria, Morro do Lena -- Alto do Vieiro, 2411-901 Leiria, Portugal}
\email{rui.santos@ipleiria.pt} 

\author{M.\ F\'atima Brilhante}
\address{Faculdade de Ci\^encias e Tecnologia, Universidade dos A\c{c}ores, Rua da M\~ae de Deus, 9500-321 Ponta Delgada, Portugal}
\email{maria.fa.brilhante@uac.pt} 


\author{Sandra Mendon\c{c}a}
\address{Departamento de Matem\'atica --- FCEE, Universidade da Madeira,  9020-105 Funchal, Portugal}
\email{sandram@staff.uma.pt} 



\thanks{The authors are members of Centro de Estat\'istica e Aplica\c{c}\~oes, Universidade de Lisboa (CEAUL). This research has been partially supported by National Funds through FCT --- Funda\c{c}\~ao para a Ci\^encia e Tecnologia, project UIDB/00006/2020 (\url{https://doi.org/10.54499/UIDB/00006/2020})}

\begin{abstract}
The classical theory for the meta-analysis of $p$-values is based on the assumption that if the overall null hypothesis is true, then all $p$-values used in a chosen combined test statistic are genuine, i.e., are observations from independent and identically distributed standard uniform random variables. However, the pressure felt by most researchers to publish,  which is worsen by publication bias, can originate fake $p$-values to be reported, usually Beta(1,2) distributed. In general, the existence of fake $p$-values in a sample of $p$-values to be combined is unknown, and  if, for some reason, there is information that they do exist, their number will most likely be unknown as well.  Moreover, even if fake $p$-values are accounted for, the cumulative distribution function of  classical combined test statistics does not have  a closed-form expression that facilitates its practical usage. To overcome this problem, tables with estimated critical values are supplied  for the commonly used combined tests for the meta-analysis of $p$-values when a few of them are fake ones, i.e., $\text{Beta}(1,2)$ distributed.
\end{abstract}

\maketitle

\section{Introduction}
The concept  of $p$-value, defined as the probability of obtaining a result equal to or more extreme than what was actually observed, under a  null hypothesis $H_0$ of no effect or of no difference, plays a key role in Fisher's  theory of significance testing. Under the assumption that $H_0$ is true, the  $p$-value is an observation from a random variable $P\sim \text{Uniform}(0,1)$, which is a simple consequence of the probability integral transform theorem. On the other hand, a $p$-value smaller than 0.05, considered by Fisher  (\citealp{Fisher22,Fisher25}) to  be a threshold for a significant value,   only  hints  that    the experiment should be  repeated; quoting \cite{Fisher26}, p.\ 85,
\begin{quotation}
"\textit{A scientific fact should be regarded as experimentally established only if a properly designed experiment rarely fails to give this {\rm [$p=0.05$]}  level of significance}",
\end{quotation}
thus raising the issue of independently replicating experiments and  of combining tests, i.e., testing the overall null  hypothesis $H_0^*:~{\rm all~ the} ~ H_{0,k},~k=1,\dotsc, n,~  {\rm are~ true}~vs.$\ the alternative hypothesis $H_A^*:~{\rm not ~all} ~ H_{A,k}$, $k=1,\dotsc, n$,  are  false.

The classical combined test statistics for the meta-analysis of $p$-values are briefly described in  Section \ref{genuine}.  Section \ref{fake}  discusses the possibility of some of the $p$-values to be combined being   fake $p$-values, due to malpractice or to fraudulent  non-reported replication of an experiment. As in such framework there is no closed-form expressions for the cumulative distribution functions of the combined test statistics, in Section \ref{simulation},  the simulation methodology used to obtain estimated high and low quantiles is described. The tables with the quantiles are supplied in the Appendix.


\section{Combined Test Statistics for the Meta-Analysis of Genuine $\boldsymbol p$-Values}\label{genuine}

Under the validity of the overall null hypothesis $H_0^*$, there are several combined methods to choose from to carry out meta-analysis of genuine $p$-values. The most commonly used combined  methods are listed below and for each one, its test statistic $T(P_1,\dotsc,P_n)$ is indicated. 
\begin{enumerate}

\item {Tippett's  method} (\citealp{Tippett}):
\[
T_T(P_1,\dotsc,P_n)= \min \{ P_1,\dotsc,P_n\} = P_{1:n} \, .
\]

\item Fisher's method (\citealp{Fisher32}):
\[
T_F(P_1,\dotsc,P_n)=-2\sum_{k=1}^n \ln P_k  \, .
\]

\item {Pearson's geometric mean method} (\citealp{Pearson33}):
\[
T_{\mathcal G_n}(P_1,\dotsc,P_n)= \left(\prod_{k=1}^n P_k\right)^{1/n} \, .
\]

\item {Pearson's minimum of geometric means method} (\citealp{Pearson34}):
\[
T_{\mathcal G_n,\mathcal G_n^*}(P_1,\dotsc,P_n)= \min \left\{\left(\prod_{k=1}^n P_k\right)^{1/n},\left(\prod_{k=1}^n (1-P_k)\right)^{1/n}\right\}\, ,
\]
which is useful  in a bilateral context.

\item {Stouffer et al.'s method} (\citealp{SSDSW}):
\[
T_S(P_1,\dotsc,P_n)=\sum_{k=1}^n \frac{\Phi^{-1}(P_k)}{\sqrt{n}} \, ,
\]
with $\Phi^{-1}$ denoting the inverse of the standard Gaussian  cumulative distribution function.

\item {Wilkinson's method} (\citealp{Wilkinson}):
\[
T_W(P_1,\dotsc,P_n)= P_{k:n}\, , \quad \text{for a } k =1,\dotsc, n,
\]
thus generalizing Tippett's method.

\item Edgington's arithmetic method (\citealp{Edgington}):
\[
T_E(P_1,\dotsc,P_n)=\overline P_n=\frac{\sum_{k=1}^n P_k}{n} \, .
\]

\item  {Mudholkar and George's method} (\citealp{MG}): 
\[
T_{MG}(P_1,\dotsc,P_n)=-\sum_{k=1}^n \ln \left(\frac{P_k}{1-P_k}\right) \, .
\]

\item Wilson's harmonic mean method (\citealp{Wilson}):
\[
  T_{\mathcal H_n}(P_1,\dots,P_n)= \frac{n}{\sum_{k=1}^n 1/P_k} \, .
 \]

\item Chen's method (\citealp{Chen}):
\[ 
T_C(P_1,\dotsc,P_n)=\sum_{k=1}^n \big[ \Phi^{-1}(P_k) \big]^2 \, .  
\]
\end{enumerate}

\cite{LF1971,LF1973}  showed   that, under mild  conditions,  Fisher's method  is optimal for combining independent tests, while \cite{MB} and \cite{Whitlock} recommend Stouffer et al.'s method. On the other hand, \cite{Birnbaum} showed that every monotone combined test procedure is admissible,  i.e.,  provides a most powerful test against some alternative hypothesis for combining some collection of tests, and   therefore is  optimal for some combined testing situation whose goal is to  harmonize possibly conflicting evidence, or to pool inconclusive evidence. Therefore,  to reach an overall decision, it is advisable to take into consideration the above list of combined tests.


\section{Fake $\boldsymbol p$-Values \label{fake}}

It is however  naive to assume that all observed $p_k$'s to be combined are genuine $p$-values. Quoting Fisher  on his critical  judgement of Mendel's work (\citealp{Fisher36}, p.\ 132),
\begin{quotation}
"\textit{[...] the data of most, if not all, of the experiments have been falsified so as to agree closely with Mendel's expectations."}
\end{quotation}
As a result, the Mendel-Fisher controversy  started a  debate on scientific research malpractices and scientific  fraudes  (ab)using  Statistics (see \citealp{Franklin}, and \citealp{PB}).

As publication bias implies that non-significant results are less likely to be published (\citealp{GST, PRVV, JZH, LC}), this fact can  lead to inappropriate replication of experiments  when the observed  $p$-value of the initial experiment is greater than 0.05,   in the hope of  obtaining a "better"\ $p$-value,  i.e., a smaller $p$-value. In many cases, what will happen is that  both $p$-values obtained are greater than 0.05, thus  not being worthy of  being reported, and  consequently the replication of the experiment is discouraged.

In view of the possibility of an experimenter reporting the smallest of two $p$-values when a second $p$-value is obtained,   the $p$-value that is actually  reported is a fake $p$-value that is Beta(1,2) distributed, since it is  the minimum of two independent standard uniform random variables. 

This clearly changes the distribution of the combined test statistics. For instance, in Fisher's method, if $P_1^*,\dotsc,P_\ell^*$ are fake $p$-values such that $P_k^* \sim \text{Beta}(1,2)$ for $k=1,\dotsc,\ell$, and $P_{\ell+1},\dotsc, P_n$ are genuine $p$-values, i.e., $P_j \sim \text{Uniform}(0,1)$ for $j=\ell+1,\dotsc,n$, then \mbox{$-4\sum_{k=1}^\ell \ln(1-P_k^*) \sim \chi_{2\ell}^2$} and \mbox{$-2\sum_{j=\ell+1}^n \ln P_j \sim \chi_{2(n-\ell)}^2$}, and therefore \mbox{$-4\sum_{k=1}^\ell \ln(1-P_k^*)-2\sum_{j=\ell+1}^n \ln P_j \sim\chi_{2n}^2$}. Moreover,  there are so far no reliable procedures  to estimate  which $p$-values are the fake ones, let alone how many exist.

With the exception of Tippett's statistic $T_T(P_1,\dotsc,P_n)=P_{1:n}$,  the other  combined test    statistics   have a distribution that does not allow  a closed-form expression for its cumulative distribution  function when there are \mbox{$n_f \ge 1$} fake $p$-values in the sample. For further details, see \cite{BGMPS}.


\section{Simulation Notes}\label{simulation}

As already mentioned, it is in general impossible to know whether there are or not fake $p$-values  among the set of  $p$-values to be combined. Therefore, a realistic approach is to examine  possible  scenarios and assess how the existence of fake $p$-values can affect the decision on the overall null hypothesis $H_0^*$.  

The combination of genuine and false $p$-values for any of the above methods requires the existence of critical values. In the collection of indicated methods, Tippett's statistic $T_T(P_1,\dotsc,P_n)=P_{1:n}$    is a rather exceptional case,  since when there are \mbox{$n_f \ge 1$} fake $p$-values in the sample,  we have $P_{1:n} \vert H_0^*~\sim~\text{Beta}(1,n+n_f)$.  However, for  the other      combined test    statistics,     there are no closed-form  expressions for the cumulative distribution functions    when there are \mbox{$n_f \ge 1$} fake $p$-values in the sample.  This   situation  makes it hard to obtain exact critical  values  that are necessary for hypothesis testing.   To overcome this problem,  a simple simulation was carried out with the software R  (version 4.3.1), a language and environment for statistical computing (\citealp{RCore}),   to   obtain the  $q$-th quantile estimate,    for $q = 0.005,0.01,0.025,0.05,0.1,0.9,0.95,0.975,0.99,0.995$,     of the combined   test statistics' distributions   when some of the $p$-values are fake ones.

Since each genuine $p$-value follows a standard  uniform distribution and a fake $p$-value a $\text{Beta}(1,2)$ distribution,  to estimate the quantiles of the  test statistics, the following methodology was  used.  

  A total of \mbox{$N=4999$}   samples  of $p$-values of  size   $n=3,\dotsc, 26$ were  generated,  for    which   there are  at most $ \lfloor n/3 \rfloor$ fake $p$-values       in each  sample  (i.e., $ n_f \le \lfloor n/3 \rfloor$).   For  each  generated  sample, the combined test  statistic $T(P_1,\dotsc,P_n)$ was computed,  thus  originating an observed   sample of the  test  statistic
$(t_{1;n,n_f},\dotsc,t_{N;n,n_f})$.    The sample of order statistics  $(t_{1:N;n,n_f},\dotsc,t_{N:N;n,n_f})$ was  then  used to estimate the $q$-th quantile  as $\widehat t_{n,n_f;q}=t_{\lfloor q(N+1)\rfloor :N;n,n_f}$ (see \citealp{Davison}, pp.\ 18-19). 
This procedure was  repeated   \mbox{$R=50$} times  so that a sample of size $R$ of estimates of each $q$-th quantile was  recorded, i.e.,
$(\widehat t_{n,n_f;q}^{(1)},\dotsc, \widehat t_{n,n_f;q}^{(R)})$. The final estimate of the $q$-th quantile of $T(P_1,\dotsc,P_n)$ is the average of the values $\widehat t_{n,n_f;q}^{(i)}$, $i=1,\dotsc,R$, i.e.,
\[
\widetilde T_{n,n_f;q} = \frac{1}{R} \sum_{i=1}^R \widehat t_{n,n_f;q}^{(i)} \,\; , {\text{with standard  error}}\,\;  s_{\widetilde T_{n,n_f;q}} =\sqrt{ \frac{\sum_{i=1}^{R} \left(\widehat{t}_{n,n_f;q}^{(i)} - \widetilde{T}_{n,n_f;q}\right)^2}{R(R-1)}}  \, .
\]

 By applying  the central limit theorem, an approximate  $(1-\alpha)100\%$ confidence interval  for the $q$-th quantile of $T(P_1,\dotsc,P_n)$  can be easily  computed using  the interval 
\begin{equation*}
\left(\widetilde{T}_{n,n_f;q} - z_{1-\alpha/2} \times s_{\widetilde{T}_{n,n_f;q}}\,,\, \widetilde{T}_{n,n_f;q} + z_{1-\alpha/2}  \times s_{\widetilde{T}_{n,n_f;q}}\right) \, ,
\end{equation*}
where $z_{q}$ denotes the $q$-th quantile of the standard Gaussian distribution.

 The tables with the  estimated quantiles, in a few cases  exact, are supplied in the  Appendix for sample sizes  $n=3,\dotsc, 26$, and when  there are  at most $n_f\leq \lfloor n/3\rfloor$ fake $p$-values. Notice that when only genuine $p$-values are in the  sample, i.e., \mbox{$n_f =0$},   the  critical values  of $T_F$, $T_S$, $T_C$,   $T_W$ and $T_{\mathcal{G}_n}$  indicated   in the tables   are  exact (the exact  critical values  of  $T_{\mathcal{G}_n}$  were  obtained  with  Mathematica v12).    This also  holds   true for the case of  the critical values of   Tippett's  combined test  statistic  \mbox{$T_T(P_1,\dotsc,P_n)=P_{1:n+n_f}\sim \text{Beta}(1,n+n_f)$},  regardless of the value of  $n_f$. We have, nonetheless,  simulated  the  critical values for Tippett's statistic,   just  for comparison purposes   as shown below.   
  
  The  estimates obtained with  the methodology described above can be compared with the exact  values to establish the reliability of the simulation results.   Figure \ref{Graf_SimN} illustrates the convergence of the estimated cumulative distribution function as a function of the number of samples. It shows   that there is  almost no difference    between the use of \mbox{$N=2500$} or \mbox{$N=4999$} samples for Tippett's   statistic $T_T(P_1,\dotsc,P_n)=P_{1:n}$, with \mbox{$n=5$} and \mbox{$n_f=3$}, and  also  for   Chen's statistic $T_{C}(P_1,\dotsc,P_n)=\sum_{k=1}^n [\Phi^{-1}(P_n)]^2$,   with $n=10$ and $n_f=0$. 
  
   \begin{figure}[!htb]
\begin{subfigure}{0.4\linewidth}
\centering
\includegraphics[scale=0.4]{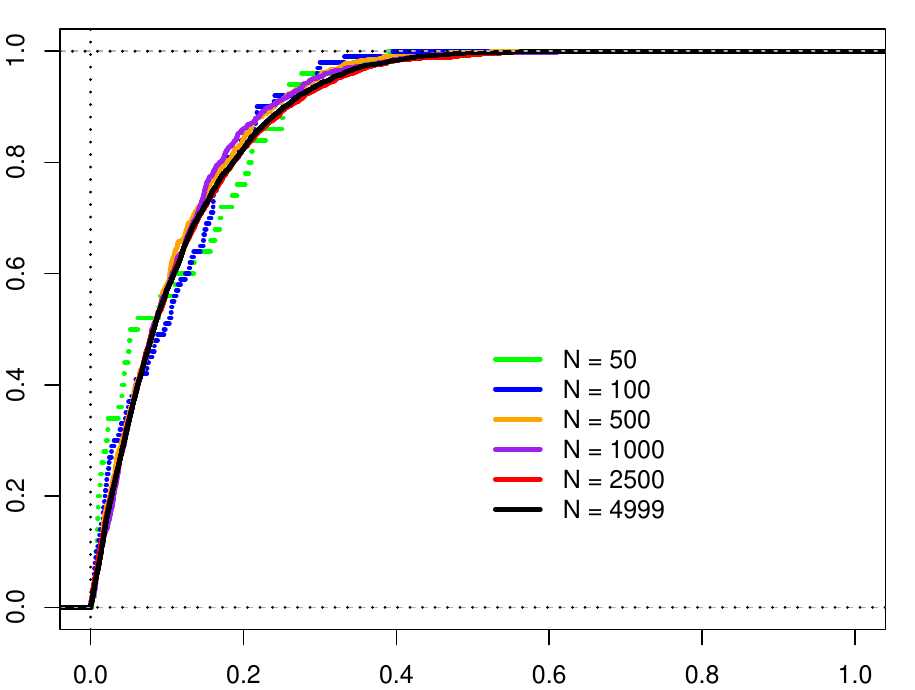}
\caption{Tippett's  method:  $n=5$ and $n_f=3$}
\end{subfigure}
\begin{subfigure}{0.4\linewidth}
\centering
\includegraphics[scale=0.4]{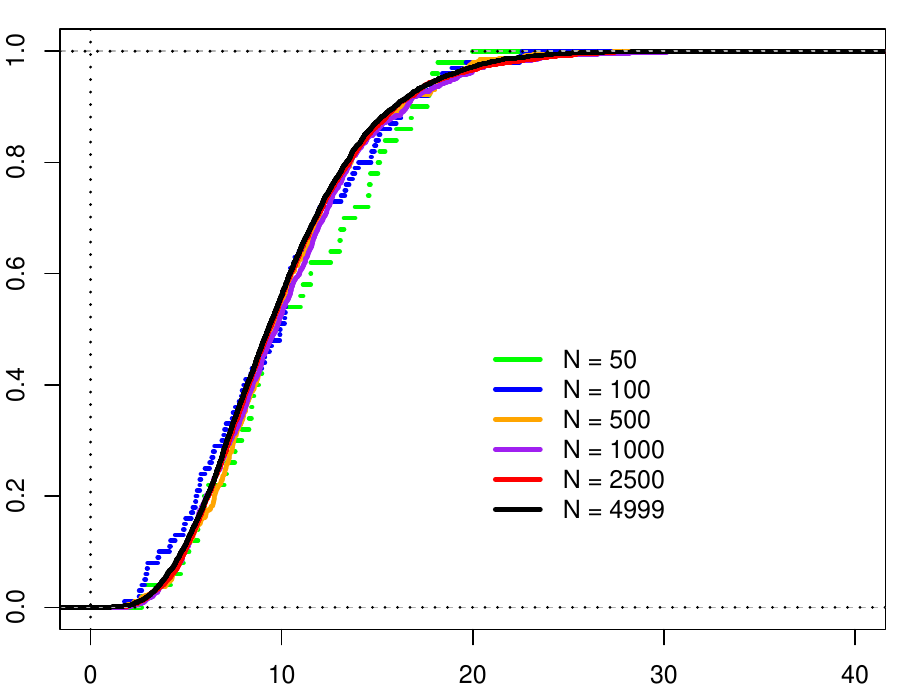}
\caption{Chen's  method: $n=10$ and $n_f=0$}
\end{subfigure}
\caption{Estimated cumulative distribution functions using  samples of size \mbox{$N=50,100,500,1000,2500,4999$}}\label{Graf_SimN} 
\end{figure}
 
 Figure \ref{Graf_Sim_rep50}  depicts  the estimated cumulative distribution function   for each of the $R=50$    replicas. It  reveals great stability   for   \mbox{$N=4999$} samples since the 50  curves   seem to overlap perfectly.     This shows  that  the methodology    used  has produced  accurate and precise estimates of the quantiles.   

\begin{figure}[!htb]
\begin{subfigure}{0.4\linewidth}
\centering
\includegraphics[scale=0.4]{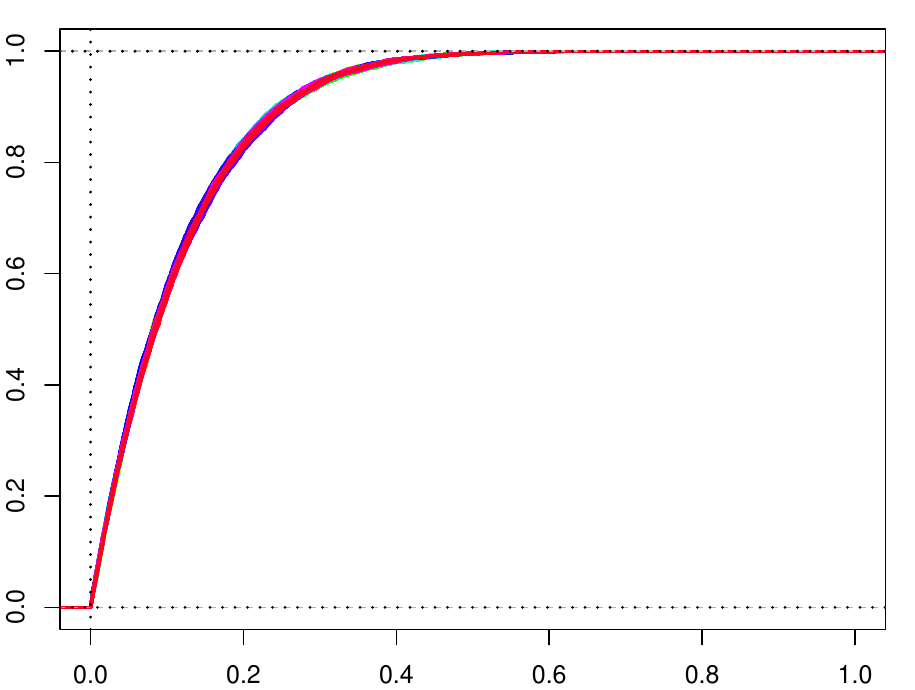}
\caption{Tippett's method: $n=5$ and $n_f=3$}
\end{subfigure}
\begin{subfigure}{0.4\linewidth}
\centering
\includegraphics[scale=0.4]{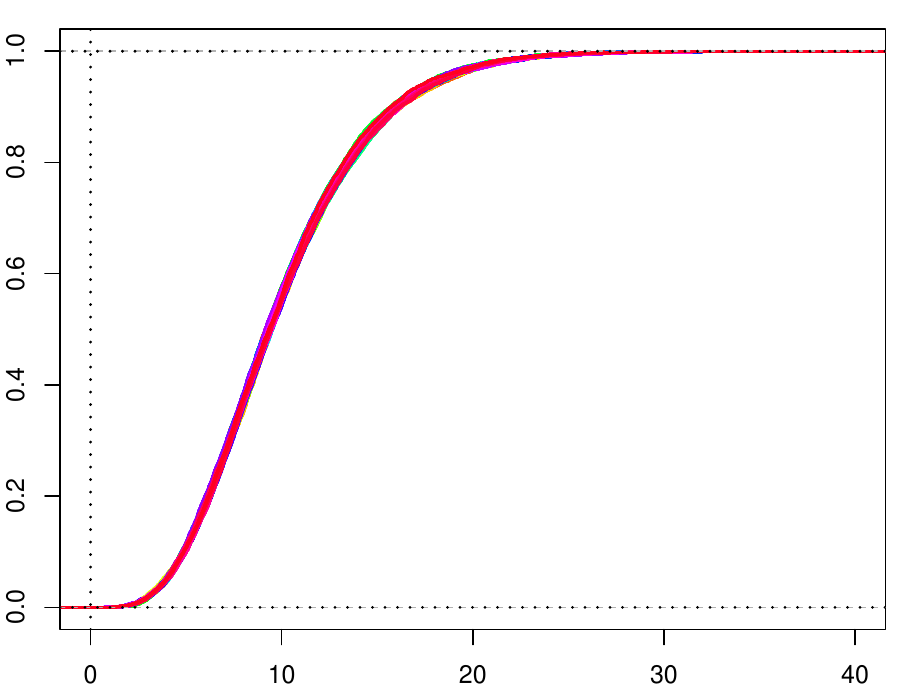}
\caption{Chen's  method: $n=10$ and $n_f =0$}
\end{subfigure}
\caption{Fifty estimated cumulative distribution function with $N=4999$ samples each}\label{Graf_Sim_rep50}  
\end{figure}

 Since  Tippett's statistic has a known distribution, namely a \text{Beta}$(1,n+n_f)$ distribution, and the same goes for   Chen's statistic, which is chi-square distributed with $n$ degrees of freedom  if  \mbox{$n_f=0$}, the estimated cumulative distribution functions of both cases  can be compared to the theoretical ones. Therefore, Figure~\ref{Graf_Sim} displays the theoretical and the estimated cumulative distribution functions   using \mbox{$N=4999$} samples, for Tippett's minimum method using \mbox{$n=5$} and \mbox{$n_f=3$} (\text{Beta}$(1,8)$ distribution), and for   Chen's  method using  $n=10$ and $n_f=0$ ($\chi_{10}^2$ distribution). 
 
 \begin{figure}[!htb]
\begin{subfigure}{0.4\linewidth}
\centering
\includegraphics[scale=0.52]{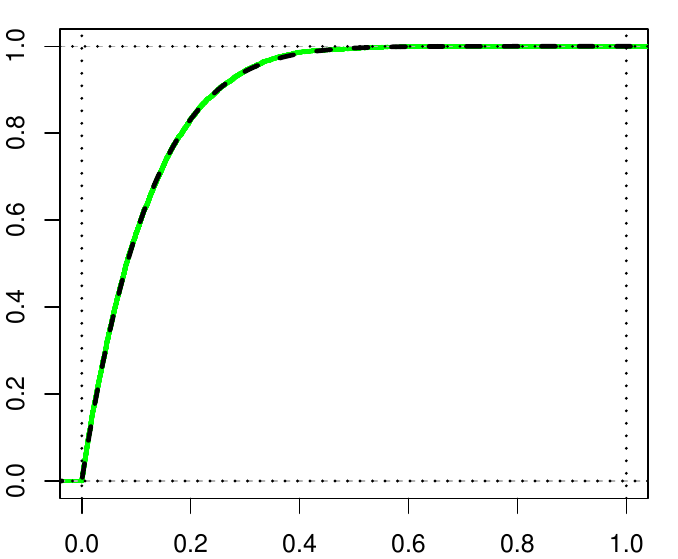}
\caption{Tippett's  method: $n=5$ and $n_f=3$}
\end{subfigure}
\begin{subfigure}{0.4\linewidth}
\centering
\includegraphics[scale=0.52]{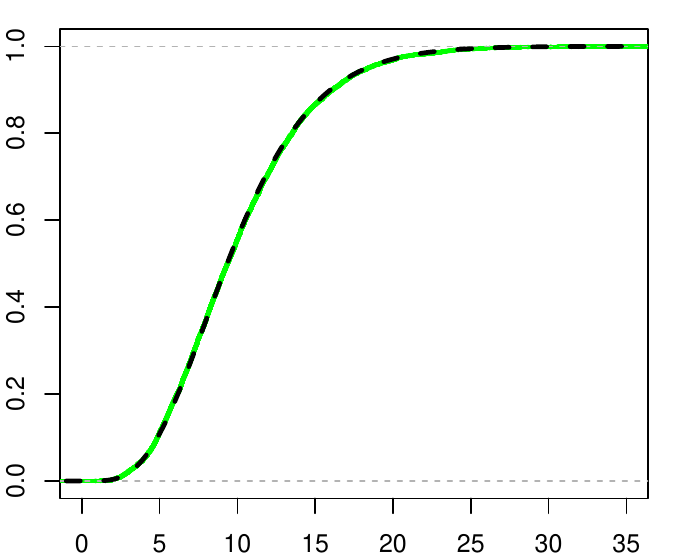}
\caption{Chen's  method: $n=10$ and $n_f=0$}
\end{subfigure}
\caption{Theoretical (black dashed line) and estimated (green solid line)  cumulative distribution functions }\label{Graf_Sim} 
\end{figure}


\section{Conclusion}

Combining genuine and fake $p$-values with any of the combined methods seen in Section~\ref{genuine}, requires the tabulation of critical values, because the cumulative distribution functions of these statistics are generally not known in a closed-form expression. This work provides estimated quantiles, and their standard errors, for the classical combined  test statistics for sample sizes  $n=3,\dotsc, 26$ when  there are at most $n_f\leq \lfloor n/3\rfloor$ fake $p$-values.
\vspace{1cm}


\appendix

\section{}

The  tables with the critical values  for the listed methods are organized as follows:

\bigskip

\begin{center}
\renewcommand{\arraystretch}{2}
\begin{tabular}{llc}
\toprule
Method & Statistic  & Table  
\\
\midrule
Fisher & $-2\sum_{k=1}^n \ln P_k$ &  \ref{Table.Fisher} 
\\
 Stouffer et al. & $\frac{\sum_{k=1}^n \Phi^{-1} (P_k)}{\sqrt{n}}$ & \ref{Table.Stouffer}
\\
Chen & $\sum_{k=1}^n \left[\Phi^{-1}(P_k)\right]^2 $ & \ref{Table.Chen}
\\
Mudholkar and George  & $- \sum_{k=1}^n \ln \!\left(\frac{P_k}{1-P_k} \right)$ & \ref{Table.MG}
\\
Geometric mean & $\left(\prod_{k=1}^n P_k\right)^{1/n} $ & \ref{Table.GM}
\\
Minimum of geometric means  & $ \min \left\{\left(\prod_{k=1}^n P_k\right)^{1/n},\left(\prod_{k=1}^n (1-P_k)\right)^{1/n}\right\}$ & \ref{Table.mGM}
\\
Wilson  & $\frac{n}{\sum_{k=1}^n 1/P_k} $ & \ref{Table.Wilson}
\\
Edgington  & $\frac{1}{n} \sum_{k=1}^n P_k $ & \ref{Table.Edgington}
\\
Tippett  & $P_{1:n}$  & \ref{Table.Tippett}
\\
Wilkinson  & $P_{n:n}$ & \ref{Table.Wilkinson}
\\
\bottomrule
\end{tabular}
\end{center}

\vspace*{2cm}
$ ~~ $

%

 
 
\section*{Fisher's statistic  $\boldsymbol{T_F(P_1,\dotsc,P_n) = -2  \sum_{k=1}^n \ln P_k} $}

\vspace{-0.4cm}
\setlength{\tabcolsep}{4pt}
\begin{table}[H]  \small
\caption{Quantiles of $T_F$ with $n_f$ fake $p$-values [estimate\,(standard error)]\label{Table.Fisher}}
\vspace{-0.4cm}
		\newcolumntype{C}{>{\centering\arraybackslash}X}
		\begin{tabularx}{\textwidth}{ccXXXXX}
		\toprule
$n$ & $n_f$ &  {0.005} & {0.010} & {0.025} & {0.050} & {0.100} 
\\
\midrule
3 & 0 & $  0.6757 $ & $  0.8721 $ & $  1.2373 $ & $  1.6354 $ & $  2.2041 $\\
3 & 1 & $ 1.1233\,(0.0102) $ & $ 1.3877\,(0.0074) $ & $ 1.8499\,(0.0069) $ & $ 2.3152\,(0.0065) $ & $ 2.9743\,(0.0078) $\\
3 & 2 & $ 1.6414\,(0.0114) $ & $ 1.9648\,(0.0108) $ & $ 2.4918\,(0.0095) $ & $ 3.0436\,(0.0094) $ & $ 3.7700\,(0.0082) $\\
3 & 3 & $ 3.0745\,(0.0200) $ & $ 3.5185\,(0.0154) $ & $ 4.2728\,(0.0153) $ & $ 5.0054\,(0.0110) $ & $ 5.9643\,(0.0081) $\\
[4pt]
4 & 0 & $  1.3444 $ & $  1.6465 $ & $  2.1797 $ & $  2.7326 $ & $  3.4895 $\\
4 & 1 & $ 1.8790\,(0.0129) $ & $ 2.2275\,(0.0119) $ & $ 2.8378\,(0.0087) $ & $ 3.4620\,(0.0073) $ & $ 4.2996\,(0.0073) $\\
4 & 2 & $ 2.4515\,(0.0144) $ & $ 2.8545\,(0.0145) $ & $ 3.5475\,(0.0128) $ & $ 4.2196\,(0.0104) $ & $ 5.1197\,(0.0104) $\\
\bottomrule
\end{tabularx}
\end{table}

\setcounter{table}{0}
			
\begin{table}[H] \small
\caption{Quantiles of $T_F$ with $n_f$ fake $p$-values (continued)}
\vspace{-0.2cm}
		\newcolumntype{C}{>{\centering\arraybackslash}X}
		\begin{tabularx}{\textwidth}{ccXXXXX}
			\toprule
$n$ & $n_f$ &  {0.900} & {0.950} & {0.975} & {0.990} & {0.995} 
\\
\midrule
3 & 0 & $ 10.6446 $ & $ 12.5916 $ & $ 14.4494 $ & $ 16.8119 $ & $ 18.5476 $\\
3 & 1 & $ 11.8574\,( 0.0188) $ & $ 13.8175\,( 0.0229) $ & $ 15.7130\,( 0.0325) $ & $ 18.0767\,( 0.0580) $ & $ 19.8817\,( 0.0727) $\\
3 & 2 & $ 13.0346\,( 0.0188) $ & $ 15.0499\,( 0.0240) $ & $ 16.8877\,( 0.0322) $ & $ 19.2477\,( 0.0490) $ & $ 21.0143\,( 0.0869) $\\
3 & 3 & $ 16.8370\,( 0.0196) $ & $ 19.0401\,( 0.0237) $ & $ 21.1389\,( 0.0305) $ & $ 23.7008\,( 0.0549) $ & $ 25.6451\,( 0.0816) $\\
[4pt]
4 & 0 & $ 13.3616 $ & $ 15.5073 $ & $ 17.5345 $ & $ 20.0902 $ & $ 21.9550 $\\
4 & 1 & $ 14.5354\,( 0.0208) $ & $ 16.7228\,( 0.0254) $ & $ 18.6995\,( 0.0366) $ & $ 21.3005\,( 0.0533) $ & $ 23.1989\,( 0.0728) $\\
4 & 2 & $ 15.7034\,( 0.0209) $ & $ 17.8778\,( 0.0250) $ & $ 19.9409\,( 0.0347) $ & $ 22.5461\,( 0.0441) $ & $ 24.3881\,( 0.0695) $\\
\bottomrule
\end{tabularx}
\end{table}

\setcounter{table}{0}

\begin{table}[H] \small
\caption{Quantiles of $T_F$ with $n_f$ fake $p$-values (continued)}
\vspace{-0.2cm} 
		\newcolumntype{C}{>{\centering\arraybackslash}X}
		\begin{tabularx}{\textwidth}{ccXXXXX}
			\toprule
${n}$ & ${n_f}$ &  {0.005} & {0.010} & {0.025} & {0.050} & {0.100} 
\\
\midrule
14 & 0 & $ 12.4613 $ & $ 13.5647 $ & $ 15.3079 $ & $ 16.9279 $ & $ 18.9392 $\\
14 & 1 & $ 13.2257\,( 0.0338) $ & $ 14.3383\,( 0.0268) $ & $ 16.0969\,( 0.0261) $ & $ 17.7791\,( 0.0274) $ & $ 19.8547\,( 0.0234) $\\
14 & 2 & $ 14.0206\,( 0.0388) $ & $ 15.1606\,( 0.0341) $ & $ 16.9626\,( 0.0281) $ & $ 18.6493\,( 0.0243) $ & $ 20.7549\,( 0.0210) $\\
14 & 3 & $ 14.7976\,( 0.0455) $ & $ 15.9491\,( 0.0332) $ & $ 17.8653\,( 0.0294) $ & $ 19.5555\,( 0.0227) $ & $ 21.6676\,( 0.0205) $\\
14 & 4 & $ 15.5441\,( 0.0440) $ & $ 16.8032\,( 0.0370) $ & $ 18.7083\,( 0.0247) $ & $ 20.4534\,( 0.0237) $ & $ 22.5978\,( 0.0192) $\\
[4pt]
15 & 0 & $ 13.7867 $ & $ 14.9535 $ & $ 16.7908 $ & $ 18.4927 $ & $ 20.5992 $\\
15 & 1 & $ 14.5767\,( 0.0346) $ & $ 15.7326\,( 0.0300) $ & $ 17.6112\,( 0.0251) $ & $ 19.3670\,( 0.0251) $ & $ 21.4954\,( 0.0222) $\\
\bottomrule
\end{tabularx}
\end{table}

\setcounter{table}{0}
			
\begin{table}[H] \small
\caption{Quantiles of $T_F$ with $n_f$ fake $p$-values (continued)}
\vspace{-0.2cm}
		\newcolumntype{C}{>{\centering\arraybackslash}X}
		\begin{tabularx}{\textwidth}{ccXXXXX}
			\toprule
${n}$ & ${n_f}$ &  {0.900} & {0.950} & {0.975} & {0.990} & {0.995} 
\\
\midrule
14 & 0 & $ 37.9159 $ & $ 41.3371 $ & $ 44.4608 $ & $ 48.2782 $ & $ 50.9934 $\\
14 & 1 & $ 39.0728\,( 0.0274) $ & $ 42.5142\,( 0.0376) $ & $ 45.6215\,( 0.0550) $ & $ 49.4272\,( 0.0795) $ & $ 52.1645\,( 0.1260) $\\
14 & 2 & $ 40.0977\,( 0.0283) $ & $ 43.5811\,( 0.0348) $ & $ 46.7159\,( 0.0459) $ & $ 50.5242\,( 0.0823) $ & $ 53.3041\,( 0.1142) $\\
14 & 3 & $ 41.1491\,( 0.0287) $ & $ 44.6187\,( 0.0355) $ & $ 47.7914\,( 0.0488) $ & $ 51.6902\,( 0.0764) $ & $ 54.5236\,( 0.1052) $\\
14 & 4 & $ 42.2346\,( 0.0299) $ & $ 45.6557\,( 0.0394) $ & $ 48.8320\,( 0.0505) $ & $ 52.8002\,( 0.0840) $ & $ 55.6239\,( 0.1122) $\\
[4pt]
15 & 0 & $ 40.2560 $ & $ 43.7730 $ & $ 46.9792 $ & $ 50.8922 $ & $ 53.6720 $\\
15 & 1 & $ 41.4265\,( 0.0271) $ & $ 44.9727\,( 0.0372) $ & $ 48.1407\,( 0.0622) $ & $ 51.9682\,( 0.0837) $ & $ 54.9381\,( 0.1320) $\\
\bottomrule
\end{tabularx}
\end{table}

\setcounter{table}{0}

\begin{table}[H] \small
\caption{Quantiles of $T_F$ with $n_f$ fake $p$-values (continued)}
\vspace{-0.2cm}
		\newcolumntype{C}{>{\centering\arraybackslash}X}
		\begin{tabularx}{\textwidth}{ccXXXXX}
			\toprule
${n}$ & ${n_f}$ &  {0.005} & {0.010} & {0.025} & {0.050} & {0.100} 
\\
\midrule
21 & 0 & $ 22.1385 $ & $ 23.6501 $ & $ 25.9987 $ & $ 28.1440 $ & $ 30.7654 $\\
21 & 1 & $ 22.9995\,( 0.0614) $ & $ 24.5831\,( 0.0506) $ & $ 26.9548\,( 0.0331) $ & $ 29.0953\,( 0.0342) $ & $ 31.7533\,( 0.0301) $\\
21 & 2 & $ 23.8560\,( 0.0542) $ & $ 25.3757\,( 0.0474) $ & $ 27.8044\,( 0.0385) $ & $ 30.0099\,( 0.0296) $ & $ 32.6766\,( 0.0337) $\\
21 & 3 & $ 24.5860\,( 0.0609) $ & $ 26.2364\,( 0.0454) $ & $ 28.6965\,( 0.0375) $ & $ 30.9006\,( 0.0336) $ & $ 33.5964\,( 0.0290) $\\
21 & 4 & $ 25.4484\,( 0.0589) $ & $ 27.0602\,( 0.0416) $ & $ 29.5193\,( 0.0332) $ & $ 31.8008\,( 0.0306) $ & $ 34.5078\,( 0.0269) $\\
21 & 5 & $ 26.3017\,( 0.0594) $ & $ 27.9334\,( 0.0450) $ & $ 30.3926\,( 0.0378) $ & $ 32.6594\,( 0.0337) $ & $ 35.4444\,( 0.0277) $\\
21 & 6 & $ 27.0992\,( 0.0559) $ & $ 28.7897\,( 0.0519) $ & $ 31.2928\,( 0.0341) $ & $ 33.5467\,( 0.0322) $ & $ 36.3369\,( 0.0267) $\\
\bottomrule
\end{tabularx}
\end{table}

\setcounter{table}{0}
			
\begin{table}[H] \small
\caption{Quantiles of $T_F$ with $n_f$ fake $p$-values (continued)}
\vspace{-0.2cm}
		\newcolumntype{C}{>{\centering\arraybackslash}X}
		\begin{tabularx}{\textwidth}{ccXXXXX}
			\toprule
${n}$ & ${n_f}$ &  {0.900} & {0.950} & {0.975} & {0.990} & {0.995} 
\\
\midrule
21 & 0 & $ 54.0902 $ & $ 58.1240 $ & $ 61.7768 $ & $ 66.2062 $ & $ 69.3360 $\\
21 & 1 & $ 55.1966\,( 0.0373) $ & $ 59.2490\,( 0.0547) $ & $ 62.8931\,( 0.0679) $ & $ 67.2912\,( 0.0888) $ & $ 70.6355\,( 0.1256) $\\
21 & 2 & $ 56.2570\,( 0.0372) $ & $ 60.2741\,( 0.0490) $ & $ 63.9116\,( 0.0615) $ & $ 68.3833\,( 0.0798) $ & $ 71.4143\,( 0.1097) $\\
21 & 3 & $ 57.3113\,( 0.0396) $ & $ 61.3660\,( 0.0509) $ & $ 65.0783\,( 0.0641) $ & $ 69.5426\,( 0.0853) $ & $ 72.6267\,( 0.1144) $\\
21 & 4 & $ 58.3635\,( 0.0378) $ & $ 62.4272\,( 0.0491) $ & $ 66.1785\,( 0.0636) $ & $ 70.6152\,( 0.0880) $ & $ 73.8209\,( 0.1203) $\\
21 & 5 & $ 59.4280\,( 0.0379) $ & $ 63.4785\,( 0.0479) $ & $ 67.2205\,( 0.0693) $ & $ 71.6942\,( 0.0975) $ & $ 74.8720\,( 0.1102) $\\
21 & 6 & $ 60.4852\,( 0.0381) $ & $ 64.6052\,( 0.0476) $ & $ 68.3494\,( 0.0684) $ & $ 72.8930\,( 0.0903) $ & $ 75.9574\,( 0.1291) $\\
\bottomrule
\end{tabularx}
\end{table}


\vspace*{-0.5cm}
\section*{Stouffer's statistic $\boldsymbol {T_S(P_1,\dotsc,P_n) =\sum_{k=1}^{n}\frac{\Phi^{-1}(P_k)}{\sqrt{n}}}$}
\vspace{-0.4cm}
\setcounter{table}{1}

\begin{table}[H] \small
\caption{Quantiles of $T_S$ with $n_f$ fake $p$-values [estimate\,(standard error)]\label{Table.Stouffer}}
\vspace{-0.4cm}
		\newcolumntype{C}{>{\centering\arraybackslash}X}
\begin{tabularx}{\textwidth}{ccXXXXX}
			\toprule
${n}$ & ${n_f}$ &  {0.005} & {0.010} & {0.025} & {0.050} & {0.100} 
\\
\midrule
3 & 0 & $-2.5758 $ & $-2.3264 $ & $-1.9600 $ & $-1.6448 $ & $-1.2815 $\\
3 & 1 & $-2.7843~(0.0116) $ & $-2.5422~(0.0075) $ & $-2.1943~(0.0045) $ & $-1.8918~(0.0043) $ & $-1.5461~(0.0035) $\\
3 & 2 & $-2.9555~(0.0102) $ & $-2.7278~(0.0061) $ & $-2.4084~(0.0045) $ & $-2.1292~(0.0039) $ & $-1.7993~(0.0032) $\\
3 & 3 & $-3.1266~(0.0083) $ & $-2.9044~(0.0059) $ & $-2.5719~(0.0039) $ & $-2.2914~(0.0035) $ & $-1.9708~(0.0032) $\\
[4pt]
4 & 0 & $-2.5758 $ & $-2.3264 $ & $-1.9600 $ & $-1.6448 $ & $-1.2815 $\\
4 & 1 & $-2.7606~(0.0094) $ & $-2.5253~(0.0081) $ & $-2.1726~(0.0050) $ & $-1.8669~(0.0038) $ & $-1.5160~(0.0034) $\\
4 & 2 & $-2.9519~(0.0082) $ & $-2.7209~(0.0064) $ & $-2.3713~(0.0048) $ & $-2.0801~(0.0038) $ & $-1.7460~(0.0033) $\\
\bottomrule
\end{tabularx}
\end{table}

\setcounter{table}{1}

\begin{table}[H] \small
\caption{Quantiles of $T_S$ with $n_f$ fake $p$-values (continued)}
\vspace{-0.2cm}
		\newcolumntype{C}{>{\centering\arraybackslash}X}
		\begin{tabularx}{\textwidth}{ccXXXXX}
			\toprule
${n}$ & ${n_f}$ &  {0.900} & {0.950} & {0.975} & {0.990} & {0.995} 
\\
\midrule

3 & 0 & $  1.2815 $ & $  1.6448 $ & $  1.9600 $ & $  2.3264 $ & $  2.5758 $\\
3 & 1 & $ 0.8893~(0.0036) $ & $ 1.2299~(0.0043) $ & $ 1.5267~(0.0052) $ & $ 1.8671~(0.0071) $ & $ 2.1117~(0.0091) $\\
3 & 2 & $ 0.4840~(0.0036) $ & $ 0.7994~(0.0042) $ & $ 1.0700~(0.0049) $ & $ 1.3894~(0.0061) $ & $ 1.6096~(0.0076) $\\
3 & 3 & $ 0.2682~(0.0033) $ & $ 0.5773~(0.0040) $ & $ 0.8394~(0.0047) $ & $ 1.1537~(0.0065) $ & $ 1.3628~(0.0085) $\\
[4pt]
4 & 0 & $  1.2815 $ & $  1.6448 $ & $  1.9600 $ & $  2.3264 $ & $  2.5758 $\\
4 & 1 & $ 0.9520~(0.0034) $ & $ 1.2956~(0.0036) $ & $ 1.5957~(0.0040) $ & $ 1.9615~(0.0064) $ & $ 2.2031~(0.0089) $\\
4 & 2 & $ 0.6163~(0.0034) $ & $ 0.9430~(0.0044) $ & $ 1.2270~(0.0050) $ & $ 1.5671~(0.0062) $ & $ 1.7967~(0.0083) $\\
\bottomrule
\end{tabularx}
\end{table}

\setcounter{table}{1}

\begin{table}[H] \small
\caption{Quantiles of $T_S$ with $n_f$ fake $p$-values (continued)}
\vspace{-0.2cm}
		\newcolumntype{C}{>{\centering\arraybackslash}X}
		\begin{tabularx}{\textwidth}{ccXXXXX}
			\toprule
${n}$ & ${n_f}$ &  {0.005} & {0.010} & {0.025} & {0.050} & {0.100} 
\\
\midrule
14 & 0 & $-2.5758 $ & $-2.3264 $ & $-1.9600 $ & $-1.6448 $ & $-1.2815 $\\
14 & 1 & $-2.6980~(0.0095) $ & $-2.4608~(0.0076) $ & $-2.0941~(0.0055) $ & $-1.7846~(0.0039) $ & $-1.4262~(0.0030) $\\
14 & 2 & $-2.8206~(0.0086) $ & $-2.5767~(0.0069) $ & $-2.2206~(0.0049) $ & $-1.9145~(0.0037) $ & $-1.5555~(0.0031) $\\
14 & 3 & $-2.9451~(0.0099) $ & $-2.7064~(0.0074) $ & $-2.3485~(0.0052) $ & $-2.0361~(0.0041) $ & $-1.6955~(0.0029) $\\
14 & 4 & $-3.0607~(0.0092) $ & $-2.8242~(0.0063) $ & $-2.4692~(0.0055) $ & $-2.1683~(0.0042) $ & $-1.8251~(0.0033) $\\
[4pt]
15 & 0 & $-2.5758 $ & $-2.3264 $ & $-1.9600 $ & $-1.6448 $ & $-1.2815 $\\
15 & 1 & $-2.6997~(0.0107) $ & $-2.4539~(0.0070) $ & $-2.0926~(0.0056) $ & $-1.7803~(0.0037) $ & $-1.4212~(0.0030) $\\
\bottomrule
\end{tabularx}
\end{table}

\setcounter{table}{1}

\begin{table}[H] \small
\caption{Quantiles of $T_S$ with $n_f$ fake $p$-values (continued)}
\vspace{-0.2cm}
		\newcolumntype{C}{>{\centering\arraybackslash}X}
		\begin{tabularx}{\textwidth}{ccXXXXX}
			\toprule
${n}$ & ${n_f}$ &  {0.900} & {0.950} & {0.975} & {0.990} & {0.995} 
\\
\midrule
14 & 0 & $  1.2815 $ & $  1.6448 $ & $  1.9600 $ & $  2.3264 $ & $  2.5758 $\\
14 & 1 & $ 1.1144~(0.0042) $ & $ 1.4786~(0.0051) $ & $ 1.7915~(0.0066) $ & $ 2.1507~(0.0081) $ & $ 2.3959~(0.0100) $\\
14 & 2 & $ 0.9487~(0.0040) $ & $ 1.3067~(0.0046) $ & $ 1.6209~(0.0061) $ & $ 1.9728~(0.0079) $ & $ 2.2031~(0.0089) $\\
14 & 3 & $ 0.7814~(0.0035) $ & $ 1.1345~(0.0041) $ & $ 1.4407~(0.0054) $ & $ 1.7908~(0.0077) $ & $ 2.0349~(0.0079) $\\
14 & 4 & $ 0.6182~(0.0038) $ & $ 0.9658~(0.0041) $ & $ 1.2572~(0.0056) $ & $ 1.6077~(0.0073) $ & $ 1.8515~(0.0102) $\\
[4pt]
15 & 0 & $  1.2815 $ & $  1.6448 $ & $  1.9600 $ & $  2.3264 $ & $  2.5758 $\\
15 & 1 & $ 1.1208~(0.0038) $ & $ 1.4801~(0.0049) $ & $ 1.7928~(0.0068) $ & $ 2.1588~(0.0073) $ & $ 2.4071~(0.0098) $\\
\bottomrule
\end{tabularx}
\end{table}

\setcounter{table}{1}

\begin{table}[H] \small
\caption{Quantiles of $T_S$ with $n_f$ fake $p$-values (continued)}
\vspace{-0.2cm}
		\newcolumntype{C}{>{\centering\arraybackslash}X}
		\begin{tabularx}{\textwidth}{ccXXXXX}
			\toprule
${n}$ & ${n_f}$ &  {0.005} & {0.010} & {0.025} & {0.050} & {0.100} 
\\
\midrule
21 & 0 & $-2.5758 $ & $-2.3264 $ & $-1.9600 $ & $-1.6448 $ & $-1.2815 $\\
21 & 1 & $-2.6845~(0.0082) $ & $-2.4394~(0.0074) $ & $-2.0712~(0.0054) $ & $-1.7602~(0.0043) $ & $-1.3985~(0.0032) $\\
21 & 2 & $-2.7918~(0.0092) $ & $-2.5333~(0.0067) $ & $-2.1749~(0.0058) $ & $-1.8680~(0.0037) $ & $-1.5129~(0.0033) $\\
21 & 3 & $-2.8949~(0.0095) $ & $-2.6445~(0.0068) $ & $-2.2848~(0.0053) $ & $-1.9827~(0.0040) $ & $-1.6242~(0.0033) $\\
21 & 4 & $-2.9877~(0.0101) $ & $-2.7505~(0.0072) $ & $-2.3962~(0.0048) $ & $-2.0881~(0.0039) $ & $-1.7392~(0.0033) $\\
21 & 5 & $-3.0805~(0.0096) $ & $-2.8467~(0.0070) $ & $-2.5012~(0.0052) $ & $-2.2017~(0.0036) $ & $-1.8517~(0.0036) $\\
21 & 6 & $-3.1995~(0.0103) $ & $-2.9602~(0.0075) $ & $-2.6144~(0.0049) $ & $-2.3131~(0.0041) $ & $-1.9678~(0.0030) $\\
\bottomrule
\end{tabularx}
\end{table}

\setcounter{table}{1}

\begin{table}[H] \small
\caption{Quantiles of $T_S$ with $n_f$ fake $p$-values (continued)}
\vspace{-0.2cm} 
		\newcolumntype{C}{>{\centering\arraybackslash}X}
		\begin{tabularx}{\textwidth}{ccXXXXX}
			\toprule
${n}$ & ${n_f}$ &  {0.900} & {0.950} & {0.975} & {0.990} & {0.995} 
\\
\midrule
21 & 0 & $  1.2815 $ & $  1.6448 $ & $  1.9600 $ & $  2.3264 $ & $  2.5758 $\\
21 & 1 & $ 1.1426~(0.0038) $ & $ 1.4998~(0.0046) $ & $ 1.8076~(0.0060) $ & $ 2.1686~(0.0073) $ & $ 2.4185~(0.0098) $\\
21 & 2 & $ 1.0093~(0.0041) $ & $ 1.3637~(0.0049) $ & $ 1.6698~(0.0062) $ & $ 2.0306~(0.0068) $ & $ 2.2802~(0.0103) $\\
21 & 3 & $ 0.8761~(0.0039) $ & $ 1.2284~(0.0044) $ & $ 1.5387~(0.0054) $ & $ 1.8945~(0.0061) $ & $ 2.1421~(0.0103) $\\
21 & 4 & $ 0.7429~(0.0038) $ & $ 1.0943~(0.0041) $ & $ 1.4002~(0.0048) $ & $ 1.7539~(0.0070) $ & $ 2.0047~(0.0089) $\\
21 & 5 & $ 0.6118~(0.0038) $ & $ 0.9577~(0.0043) $ & $ 1.2638~(0.0054) $ & $ 1.6125~(0.0075) $ & $ 1.8508~(0.0086) $\\
21 & 6 & $ 0.4791~(0.0034) $ & $ 0.8266~(0.0044) $ & $ 1.1287~(0.0053) $ & $ 1.4760~(0.0072) $ & $ 1.7041~(0.0103) $\\
\bottomrule
\end{tabularx}
\end{table}


\vspace*{-0.5cm}

\section*{Chen's statistic $\boldsymbol{T_{C}(P_1,\dotsc,P_n) =  \sum_{k=1}^n \left[\Phi^{-1}(P_k)\right]^2}$ }
\vspace{-0.4cm}
\setcounter{table}{2}

\begin{table}[H] \small
\caption{Quantiles of $T_{C}$ with $n_f$ fake $p$-values [estimate\,(standard error)]}\label{Table.Chen}
\vspace{-0.4cm}
		\newcolumntype{C}{>{\centering\arraybackslash}X}
\begin{tabularx}{\textwidth}{ccXXXXX}
			\toprule
${n}$ & ${n_f}$ &  {0.005} & {0.010} & {0.025} & {0.050} & {0.100} 
\\
\midrule
3 & 0 & $ 0.0717 $ & $  0.1148 $ & $  0.2158 $ & $  0.3518 $ & $  0.5844 $\\
3 & 1 & $ 0.0714\,(0.0015) $ & $ 0.1141\,(0.0017) $ & $ 0.2156\,(0.0020) $ & $ 0.3529\,(0.0020) $ & $ 0.5839\,(0.0026) $\\
3 & 2 & $ 0.0705\,(0.0013) $ & $ 0.1137\,(0.0014) $ & $ 0.2136\,(0.0017) $ & $ 0.3504\,(0.0019) $ & $ 0.5801\,(0.0026) $\\
3 & 3 & $ 0.2016\,(0.0032) $ & $ 0.2929\,(0.0031) $ & $ 0.4828\,(0.0027) $ & $ 0.7052\,(0.0031) $ & $ 1.0616\,(0.0034) $\\
[4pt]
4 & 0 & $  0.2070 $ & $  0.2971 $ & $  0.4844 $ & $  0.7107 $ & $  1.0636 $\\
4 & 1 & $ 0.2041\,(0.0027) $ & $ 0.2922\,(0.0029) $ & $ 0.4817\,(0.0028) $ & $ 0.7093\,(0.0030) $ & $ 1.0623\,(0.0034) $\\
4 & 2 & $ 0.2058\,(0.0034) $ & $ 0.2953\,(0.0034) $ & $ 0.4822\,(0.0031) $ & $ 0.7044\,(0.0037) $ & $ 1.0678\,(0.0039) $\\
\bottomrule
\end{tabularx}
\end{table}

\setcounter{table}{2}

\begin{table}[H] \small
\caption{Quantiles of $T_{C}$ with $n_f$ fake $p$-values (continued)}
\vspace{-0.2cm} 
		\newcolumntype{C}{>{\centering\arraybackslash}X}
		\begin{tabularx}{\textwidth}{ccXXXXX}
			\toprule
${n}$ & ${n_f}$ &  {0.900} & {0.950} & {0.975} & {0.990} & {0.995} 
\\
\midrule
3 & 0 & $  6.2514 $ & $  7.8147 $ & $  9.3484 $ & $ 11.3449 $ & $ 12.8382 $\\
3 & 1 & $ 6.2712\,(0.0117) $ & $ 7.8110\,(0.0151) $ & $ 9.3413\,(0.0239) $ & $ 11.3059\,( 0.0456) $ & $ 12.9217\,( 0.0556) $\\
3 & 2 & $ 6.2611\,(0.0116) $ & $ 7.8133\,(0.0192) $ & $ 9.3284\,(0.0261) $ & $ 11.3334\,( 0.0457) $ & $ 12.8526\,( 0.0608) $\\
3 & 3 & $ 7.7737\,(0.0142) $ & $ 9.4900\,(0.0164) $ & $ 11.1064\,( 0.0273) $ & $ 13.2860\,( 0.0465) $ & $ 14.8970\,( 0.0738) $\\
[4pt]
4 & 0 & $  7.7794 $ & $  9.4877 $ & $ 11.1433 $ & $ 13.2767 $ & $ 14.8603 $\\
4 & 1 & $ 7.7800\,(0.0124) $ & $ 9.4654\,(0.0201) $ & $ 11.0990\,( 0.0315) $ & $ 13.3081\,( 0.0484) $ & $ 14.9048\,( 0.0689) $\\
4 & 2 & $ 7.7842\,(0.0132) $ & $ 9.5021\,(0.0211) $ & $ 11.1261\,( 0.0259) $ & $ 13.2874\,( 0.0428) $ & $ 14.8894\,( 0.0662) $\\
\bottomrule
\end{tabularx}
\end{table}

\setcounter{table}{2}

\begin{table}[H] \small
\caption{Quantiles of $T_{C}$ with $n_f$ fake $p$-values (continued)}
\vspace{-0.2cm}
		\newcolumntype{C}{>{\centering\arraybackslash}X}
		\begin{tabularx}{\textwidth}{ccXXXXX}
			\toprule
${n}$ & ${n_f}$ &  {0.005} & {0.010} & {0.025} & {0.050} & {0.100} 
\\
\midrule
14 & 0 & $  4.0747 $ & $  4.6604 $ & $  5.6287 $ & $  6.5706 $ & $  7.7895 $\\
14 & 1 & $ 4.0575\,(0.0203) $ & $ 4.6389\,(0.0154) $ & $ 5.6200\,(0.0135) $ & $ 6.5634\,(0.0141) $ & $ 7.7778\,(0.0132) $\\
14 & 2 & $ 4.0510\,(0.0186) $ & $ 4.6349\,(0.0144) $ & $ 5.6184\,(0.0129) $ & $ 6.5607\,(0.0146) $ & $ 7.7770\,(0.0127) $\\
14 & 3 & $ 4.0332\,(0.0198) $ & $ 4.6280\,(0.0160) $ & $ 5.6139\,(0.0124) $ & $ 6.5473\,(0.0121) $ & $ 7.7885\,(0.0119) $\\
14 & 4 & $ 4.0760\,(0.0185) $ & $ 4.6329\,(0.0172) $ & $ 5.6121\,(0.0156) $ & $ 6.5535\,(0.0119) $ & $ 7.7959\,(0.0133) $\\
[4pt]
15 & 0 & $  4.6009 $ & $  5.2294 $ & $  6.2621 $ & $  7.2609 $ & $  8.5468 $\\
15 & 1 & $ 4.5751\,(0.0195) $ & $ 5.2161\,(0.0188) $ & $ 6.2502\,(0.0153) $ & $ 7.2424\,(0.0159) $ & $ 8.5373\,(0.0135) $\\
\bottomrule
\end{tabularx}
\end{table}

\setcounter{table}{2}

\begin{table}[H] \small
\caption{Quantiles of $T_{C}$ with $n_f$ fake $p$-values (continued)}
\vspace{-0.2cm} 
		\newcolumntype{C}{>{\centering\arraybackslash}X}
		\begin{tabularx}{\textwidth}{ccXXXXX}
			\toprule
${n}$ & ${n_f}$ &  {0.900} & {0.950} & {0.975} & {0.990} & {0.995} 
\\
\midrule
14 & 0 & $ 21.0641 $ & $ 23.6848 $ & $ 26.1190 $ & $ 29.1412 $ & $ 31.3194 $\\
14 & 1 & $ 21.0930\,( 0.0241) $ & $ 23.6906\,( 0.0299) $ & $ 26.1207\,( 0.0432) $ & $ 29.2008\,( 0.0608) $ & $ 31.3324\,( 0.0834) $\\
14 & 2 & $ 21.0526\,( 0.0212) $ & $ 23.6747\,( 0.0297) $ & $ 26.1357\,( 0.0424) $ & $ 29.1712\,( 0.0541) $ & $ 31.2991\,( 0.0790) $\\
14 & 3 & $ 21.0408\,( 0.0213) $ & $ 23.6740\,( 0.0321) $ & $ 26.1913\,( 0.0345) $ & $ 29.1214\,( 0.0594) $ & $ 31.3395\,( 0.0735) $\\
14 & 4 & $ 21.0319\,( 0.0255) $ & $ 23.6456\,( 0.0247) $ & $ 26.1187\,( 0.0328) $ & $ 29.0766\,( 0.0564) $ & $ 31.3618\,( 0.0893) $\\
[4pt]
15 & 0 & $ 22.3071 $ & $ 24.9958 $ & $ 27.4884 $ & $ 30.5779 $ & $ 32.8013 $\\
15 & 1 & $ 22.3282\,( 0.0246) $ & $ 25.0363\,( 0.0305) $ & $ 27.4901\,( 0.0430) $ & $ 30.6236\,( 0.0640) $ & $ 32.7754\,( 0.0763) $\\
\bottomrule
\end{tabularx}
\end{table}

\setcounter{table}{2}

\begin{table}[H] \small
\caption{Quantiles of $T_{C}$ with $n_f$ fake $p$-values (continued)}
\vspace{-0.2cm} 
		\newcolumntype{C}{>{\centering\arraybackslash}X}
		\begin{tabularx}{\textwidth}{ccXXXXX}
			\toprule
${n}$ & ${n_f}$ &  {0.005} & {0.010} & {0.025} & {0.050} & {0.100} 
\\
\midrule
21 & 0 & $  8.0336 $ & $  8.8972 $ & $ 10.2829 $ & $ 11.5913 $ & $ 13.2396 $\\
21 & 1 & $ 7.9924\,(0.0315) $ & $ 8.8641\,(0.0245) $ & $ 10.2415\,( 0.0196) $ & $ 11.5633\,( 0.0184) $ & $ 13.2188\,( 0.0157) $\\
21 & 2 & $ 7.9841\,(0.0289) $ & $ 8.8373\,(0.0239) $ & $ 10.2424\,( 0.0161) $ & $ 11.5753\,( 0.0187) $ & $ 13.2133\,( 0.0173) $\\
21 & 3 & $ 7.9340\,(0.0327) $ & $ 8.8433\,(0.0265) $ & $ 10.2615\,( 0.0192) $ & $ 11.5793\,( 0.0169) $ & $ 13.2432\,( 0.0181) $\\
21 & 4 & $ 7.9530\,(0.0317) $ & $ 8.8285\,(0.0257) $ & $ 10.2561\,( 0.0204) $ & $ 11.5576\,( 0.0152) $ & $ 13.2314\,( 0.0163) $\\
21 & 5 & $ 7.9652\,(0.0277) $ & $ 8.8569\,(0.0254) $ & $ 10.2555\,( 0.0257) $ & $ 11.5422\,( 0.0204) $ & $ 13.2282\,( 0.0179) $\\
21 & 6 & $ 7.9529\,(0.0327) $ & $ 8.8280\,(0.0297) $ & $ 10.2591\,( 0.0239) $ & $ 11.5657\,( 0.0192) $ & $ 13.2226\,( 0.0180) $\\
\bottomrule
\end{tabularx}
\end{table}

\setcounter{table}{2}

\begin{table}[H] \small
\caption{Quantiles of $T_{C}$ with $n_f$ fake $p$-values (continued)}
\vspace{-0.2cm} 
		\newcolumntype{C}{>{\centering\arraybackslash}X}
		\begin{tabularx}{\textwidth}{ccXXXXX}
			\toprule
${n}$ & ${n_f}$ &  {0.900} & {0.950} & {0.975} & {0.990} & {0.995} 
\\
\midrule
21 & 0 & $ 29.6151 $ & $ 32.6706 $ & $ 35.4789 $ & $ 38.9322 $ & $ 41.4011 $\\
21 & 1 & $ 29.6224\,( 0.0290) $ & $ 32.6951\,( 0.0316) $ & $ 35.4949\,( 0.0406) $ & $ 38.9362\,( 0.0639) $ & $ 41.4741\,( 0.0924) $\\
21 & 2 & $ 29.6071\,( 0.0308) $ & $ 32.6702\,( 0.0358) $ & $ 35.4900\,( 0.0461) $ & $ 38.9540\,( 0.0709) $ & $ 41.4047\,( 0.0942) $\\
21 & 3 & $ 29.6023\,( 0.0302) $ & $ 32.6794\,( 0.0329) $ & $ 35.4622\,( 0.0426) $ & $ 39.0220\,( 0.0576) $ & $ 41.4757\,( 0.0890) $\\
21 & 4 & $ 29.5764\,( 0.0293) $ & $ 32.6556\,( 0.0327) $ & $ 35.4171\,( 0.0423) $ & $ 38.9437\,( 0.0669) $ & $ 41.5088\,( 0.0890) $\\
21 & 5 & $ 29.5903\,( 0.0289) $ & $ 32.6191\,( 0.0381) $ & $ 35.4297\,( 0.0479) $ & $ 38.9216\,( 0.0676) $ & $ 41.4534\,( 0.1005) $\\
21 & 6 & $ 29.5769\,( 0.0286) $ & $ 32.6623\,( 0.0354) $ & $ 35.4461\,( 0.0552) $ & $ 38.9306\,( 0.0851) $ & $ 41.5634\,( 0.1003) $\\
\bottomrule
\end{tabularx}
\end{table}


\vspace*{-0.5cm}
\setcounter{table}{3}

\section*{Mudholkar and George's statistic $\boldsymbol {T_{MG}(P_1,\dotsc,P_n) = - \sum_{k=1}^n   \ln  \Big(\frac{P_k}{\,1-P_{k}}\Big)} $ }
\vspace{-0.4cm}

\begin{table}[H] \small
\caption{Quantiles of $T_{MG}$ with $n_f$ fake $p$-values [estimate\,(standard error)]}\label{Table.MG}
\vspace{-0.4cm} 
		\newcolumntype{C}{>{\centering\arraybackslash}X}
\begin{tabularx}{\textwidth}{ccXXXXX}
			\toprule
${n}$ & ${n_f}$ &  {0.005} & {0.010} & {0.025} & {0.050} & {0.100} 
\\
\midrule
3 & 0 & $ -8.6198~(0.0393) $ & $ -7.6134~(0.0301) $ & $ -6.2480~(0.0165) $ & $ -5.1533~(0.0149) $ & $ -3.9416~(0.0118) $\\
3 & 1 & $ -6.9574~(0.0350) $ & $ -6.0712~(0.0275) $ & $ -4.8361~(0.0173) $ & $ -3.8282~(0.0134) $ & $ -2.7245~(0.0115) $\\
3 & 2 & $ -5.1856~(0.0281) $ & $ -4.4038~(0.0211) $ & $ -3.3384~(0.0154) $ & $ -2.4501~(0.0124) $ & $ -1.4736~(0.0112) $\\
3 & 3 & $ -4.9959~(0.0323) $ & $ -4.1664~(0.0238) $ & $ -3.0072~(0.0171) $ & $ -2.0541~(0.0144) $ & $ -0.9579~(0.0116) $\\
[4pt]
4 & 0 & $ -9.7911~(0.0434) $ & $ -8.6937~(0.0305) $ & $ -7.1869~(0.0188) $ & $ -5.9633~(0.0166) $ & $ -4.5819~(0.0117) $\\
4 & 1 & $ -8.2969~(0.0400) $ & $ -7.2640~(0.0261) $ & $ -5.8371~(0.0162) $ & $ -4.6776~(0.0133) $ & $ -3.3928~(0.0120) $\\
4 & 2 & $ -6.7491~(0.0368) $ & $ -5.8145~(0.0253) $ & $ -4.4576~(0.0189) $ & $ -3.3986~(0.0167) $ & $ -2.2002~(0.0121) $\\
\bottomrule
\end{tabularx}
\end{table}

\setcounter{table}{3}

\begin{table}[H] \small
\caption{Quantiles of $T_{MG}$ with $n_f$ fake $p$-values (continued)}
\vspace{-0.2cm}
		\newcolumntype{C}{>{\centering\arraybackslash}X}
		\begin{tabularx}{\textwidth}{ccXXXXX}
			\toprule
${n}$ & ${n_f}$ &  {0.900} & {0.950} & {0.975} & {0.990} & {0.995} 
\\
\midrule
3 & 0 & $ 3.9576~(0.0104) $ & $ 5.1527~(0.0142) $ & $ 6.2361~(0.0185) $ & $ 7.5740~(0.0285) $ & $ 8.4832~(0.0341) $\\
3 & 1 & $ 4.7861~(0.0115) $ & $ 5.9521~(0.0156) $ & $ 7.0297~(0.0164) $ & $ 8.3501~(0.0310) $ & $ 9.3211~(0.0407) $\\
3 & 2 & $ 5.6008~(0.0103) $ & $ 6.7515~(0.0127) $ & $ 7.7873~(0.0174) $ & $ 9.0466~(0.0233) $ & $ 9.9791~(0.0452) $\\
3 & 3 & $ 7.0958~(0.0126) $ & $ 8.3824~(0.0142) $ & $ 9.5724~(0.0168) $ & $ 10.9980~(0.0269) $ & $ 12.0068~(0.0374) $\\
[4pt]
4 & 0 & $ 4.5877~(0.0111) $ & $ 5.9603~(0.0119) $ & $ 7.1707~(0.0170) $ & $ 8.6767~(0.0319) $ & $ 9.7189~(0.0449) $\\
4 & 1 & $ 5.4366~(0.0138) $ & $ 6.7650~(0.0151) $ & $ 7.9805~(0.0196) $ & $ 9.4611~(0.0293) $ & $ 10.4929~(0.0413) $\\
4 & 2 & $ 6.2678~(0.0126) $ & $ 7.5669~(0.0155) $ & $ 8.7677~(0.0202) $ & $ 10.2412~(0.0249) $ & $ 11.2712~(0.0380) $\\
\bottomrule
\end{tabularx}
\end{table}

\setcounter{table}{3}

\begin{table}[H]\small
\caption{Quantiles of $T_{MG}$ with $n_f$ fake $p$-values (continued)}
\vspace{-0.2cm}
		\newcolumntype{C}{>{\centering\arraybackslash}X}
		\begin{tabularx}{\textwidth}{ccXXXXX}
			\toprule
${n}$ & ${n_f}$ &  {0.005} & {0.010} & {0.025} & {0.050} & {0.100} 
\\
\midrule
14 & 0 & $ -17.738~(0.0768) $ & $ -15.923~(0.0581) $ & $ -13.335~(0.0462) $ & $ -11.145~(0.0346) $ & $ -8.6417~(0.0249) $\\
14 & 1 & $ -16.519~(0.0767) $ & $ -14.728~(0.0609) $ & $ -12.214~(0.0440) $ & $ -10.036~(0.0325) $ & $ -7.5628~(0.0274) $\\
14 & 2 & $ -15.178~(0.0786) $ & $ -13.544~(0.0521) $ & $ -11.078~(0.0411) $ & $ -8.9015~(0.0303) $ & $ -6.4493~(0.0269) $\\
14 & 3 & $ -13.993~(0.0588) $ & $ -12.335~(0.0557) $ & $ -9.8465~(0.0385) $ & $ -7.7456~(0.0299) $ & $ -5.3369~(0.0221) $\\
14 & 4 & $ -12.769~(0.0768) $ & $ -11.085~(0.0483) $ & $ -8.6164~(0.0362) $ & $ -6.6136~(0.0285) $ & $ -4.2494~(0.0244) $\\
[4pt]
15 & 0 & $ -18.274~(0.0816) $ & $ -16.374~(0.0637) $ & $ -13.745~(0.0470) $ & $ -11.516~(0.0302) $ & $ -8.9390~(0.0236) $\\
15 & 1 & $ -17.153~(0.0733) $ & $ -15.295~(0.0599) $ & $ -12.642~(0.0466) $ & $ -10.411~(0.0338) $ & $ -7.8590~(0.0274) $\\
\bottomrule
\end{tabularx}
\end{table}

\setcounter{table}{3}

\begin{table}[H] \small
\caption{Quantiles of $T_{MG}$ with $n_f$ fake $p$-values (continued)}
\vspace{-0.2cm}
		\newcolumntype{C}{>{\centering\arraybackslash}X}
		\begin{tabularx}{\textwidth}{ccXXXXX}
			\toprule
${n}$ & ${n_f}$ &  {0.900} & {0.950} & {0.975} & {0.990} & {0.995} 
\\
\midrule
14 & 0 & $ 8.7032~(0.0197) $ & $ 11.2050~(0.0274) $ & $ 13.3809~(0.0342) $ & $ 15.9675~(0.0521) $ & $ 17.7386~(0.0783) $\\
14 & 1 & $ 9.6136~(0.0192) $ & $ 12.0735~(0.0254) $ & $ 14.2674~(0.0354) $ & $ 16.8410~(0.0529) $ & $ 18.5527~(0.0676) $\\
14 & 2 & $ 10.4848~(0.0190) $ & $ 12.9597~(0.0246) $ & $ 15.1085~(0.0356) $ & $ 17.6650~(0.0532) $ & $ 19.4270~(0.0714) $\\
14 & 3 & $ 11.4256~(0.0193) $ & $ 13.7953~(0.0291) $ & $ 15.9788~(0.0310) $ & $ 18.5794~(0.0570) $ & $ 20.2413~(0.0687) $\\
14 & 4 & $ 12.2744~(0.0232) $ & $ 14.6462~(0.0293) $ & $ 16.7755~(0.0369) $ & $ 19.3717~(0.0475) $ & $ 21.1134~(0.0657) $\\
[4pt]
15 & 0 & $  9.0154~(0.0228) $ & $ 11.5841~(0.0283) $ & $ 13.8870~(0.0391) $ & $ 16.5466~(0.0604) $ & $ 18.4182~(0.0782) $\\
15 & 1 & $  9.9294~(0.0206) $ & $ 12.4699~(0.0272) $ & $ 14.7034~(0.0382) $ & $ 17.3382~(0.0550) $ & $ 19.1899~(0.0727) $\\
\bottomrule
\end{tabularx}
\end{table}

\setcounter{table}{3}

\begin{table}[H]\small
\caption{Quantiles of $T_{MG}$ with $n_f$ fake $p$-values (continued)}
\vspace{-0.2cm}
		\newcolumntype{C}{>{\centering\arraybackslash}X}
		\begin{tabularx}{\textwidth}{ccXXXXX}
			\toprule
${n}$ & ${n_f}$ &  {0.005} & {0.010} & {0.025} & {0.050} & {0.100} 
\\
\midrule
21 & 0 & $ -21.461~(0.0856) $ & $ -19.365~(0.0568) $ & $ -16.232~(0.0512) $ & $ -13.575~(0.0414) $ & $ -10.561~(0.0339) $\\
21 & 1 & $ -20.240~(0.0779) $ & $ -18.144~(0.0588) $ & $ -15.087~(0.0518) $ & $ -12.475~(0.0411) $ & $ -9.4884~(0.0317) $\\
21 & 2 & $ -19.128~(0.0840) $ & $ -17.022~(0.0610) $ & $ -13.942~(0.0508) $ & $ -11.356~(0.0428) $ & $ -8.3938~(0.0343) $\\
21 & 3 & $ -17.999~(0.0827) $ & $ -15.858~(0.0575) $ & $ -12.862~(0.0449) $ & $ -10.250~(0.0359) $ & $ -7.3305~(0.0338) $\\
21 & 4 & $ -16.826~(0.0753) $ & $ -14.687~(0.0616) $ & $ -11.759~(0.0442) $ & $ -9.1712~(0.0360) $ & $ -6.2553~(0.0312) $\\
21 & 5 & $ -15.604~(0.0773) $ & $ -13.565~(0.0659) $ & $ -10.642~(0.0498) $ & $ -8.0621~(0.0347) $ & $ -5.1621~(0.0314) $\\
21 & 6 & $ -14.358~(0.0830) $ & $ -12.419~(0.0619) $ & $ -9.5040~(0.0471) $ & $ -6.9743~(0.0363) $ & $ -4.0947~(0.0282) $\\
\bottomrule
\end{tabularx}
\end{table}

\setcounter{table}{3}

\begin{table}[H] \small
\caption{Quantiles of $T_{MG}$ with $n_f$ fake $p$-values (continued)}
 \vspace{-0.2cm}
		\newcolumntype{C}{>{\centering\arraybackslash}X}
		\begin{tabularx}{\textwidth}{ccXXXXX}
			\toprule
${n}$ & ${n_f}$ &  {0.900} & {0.950} & {0.975} & {0.990} & {0.995} 
\\
\midrule
21 & 0 & $ 10.6431~(0.0281) $ & $ 13.6894~(0.0353) $ & $ 16.3113~(0.0448) $ & $ 19.4303~(0.0669) $ & $ 21.6044~(0.0884) $\\
21 & 1 & $ 11.5636~(0.0281) $ & $ 14.5828~(0.0353) $ & $ 17.2626~(0.0491) $ & $ 20.3736~(0.0617) $ & $ 22.5088~(0.0780) $\\
21 & 2 & $ 12.5079~(0.0303) $ & $ 15.5019~(0.0308) $ & $ 18.1168~(0.0518) $ & $ 21.1740~(0.0607) $ & $ 23.3995~(0.0796) $\\
21 & 3 & $ 13.4081~(0.0276) $ & $ 16.4268~(0.0357) $ & $ 18.9936~(0.0456) $ & $ 22.1451~(0.0626) $ & $ 24.1913~(0.0838) $\\
21 & 4 & $ 14.3405~(0.0278) $ & $ 17.2837~(0.0355) $ & $ 19.9052~(0.0415) $ & $ 22.9388~(0.0645) $ & $ 25.0638~(0.0897) $\\
21 & 5 & $ 15.2760~(0.0285) $ & $ 18.2100~(0.0332) $ & $ 20.7823~(0.0418) $ & $ 23.7436~(0.0618) $ & $ 25.8093~(0.0833) $\\
21 & 6 & $ 16.2158~(0.0251) $ & $ 19.1302~(0.0323) $ & $ 21.6947~(0.0380) $ & $ 24.7222~(0.0656) $ & $ 26.8422~(0.0824) $\\
\bottomrule
\end{tabularx}
\end{table}



\vspace*{-0.5cm}
\section*{Geometric mean statistic $\boldsymbol{T_{{\mathcal G}_n}(P_1,\dotsc,P_n) = \Big(\prod_{k=1}^n P_k\Big)^{1/n}}$ }
\vspace*{-0.4cm}

\setcounter{table}{4}

\begin{table}[H]\small
\caption{Quantiles of $T_{{\mathcal G}_n}$ with $n_f$ fake $p$-values [estimate\,(standard error)]}\label{Table.GM}
\vspace{-0.4cm} 
		\newcolumntype{C}{>{\centering\arraybackslash}X}
\begin{tabularx}{\textwidth}{ccXXXXX}
			\toprule
${n}$ & ${n_f}$ &  {0.005} & {0.010} & {0.025} & {0.050} & {0.100} 
\\
\midrule
3 & 0 & $ 0.04544 $ & $ 0.06069 $ & $ 0.08997 $ & $ 0.12263 $ & $ 0.16963 $\\
3 & 1 & $ 0.03651\,(0.00044) $ & $ 0.04927\,(0.00047) $ & $ 0.07294\,(0.00039) $ & $ 0.10000\,(0.00038) $ & $ 0.13862\,(0.00043) $\\
3 & 2 & $ 0.03028\,(0.00043) $ & $ 0.04051\,(0.00033) $ & $ 0.05997\,(0.00032) $ & $ 0.08144\,(0.00032) $ & $ 0.11393\,(0.00036) $\\
3 & 3 & $ 0.04064\,(0.00041) $ & $ 0.05174\,(0.00036) $ & $ 0.07122\,(0.00027) $ & $ 0.09257\,(0.00027) $ & $ 0.12191\,(0.00030) $\\
[4pt]
4 & 0 & $ 0.06429 $ & $ 0.08116 $ & $ 0.11171 $ & $ 0.14393 $ & $ 0.18821 $\\
4 & 1 & $ 0.05514\,(0.00050) $ & $ 0.06984\,(0.00046) $ & $ 0.09663\,(0.00045) $ & $ 0.12367\,(0.00039) $ & $ 0.16255\,(0.00042) $\\
4 & 2 & $ 0.04752\,(0.00041) $ & $ 0.05975\,(0.00033) $ & $ 0.08273\,(0.00036) $ & $ 0.10705\,(0.00033) $ & $ 0.14047\,(0.00037) $\\
\bottomrule
\end{tabularx}
\end{table}

\setcounter{table}{4}

\begin{table}[H] \small
\caption{Quantiles of $T_{{\mathcal G}_n}$ with $n_f$ fake $p$-values (continued)}
\vspace{-0.2cm}
		\newcolumntype{C}{>{\centering\arraybackslash}X}
		\begin{tabularx}{\textwidth}{ccXXXXX}
			\toprule
${n}$ & ${n_f}$ &  {0.900} & {0.950} & {0.975} & {0.990} & {0.995} 
\\
\midrule
3 & 0 & $ 0.69256 $ & $ 0.76142 $ & $ 0.81365 $ & $ 0.86472 $ & $ 0.89349 $\\
3 & 1 & $ 0.60916\,(0.00079) $ & $ 0.67988\,(0.00074) $ & $ 0.73471\,(0.00084) $ & $ 0.79354\,(0.00098) $ & $ 0.82932\,(0.00141) $\\
3 & 2 & $ 0.53350\,(0.00073) $ & $ 0.60217\,(0.00095) $ & $ 0.66019\,(0.00104) $ & $ 0.72081\,(0.00130) $ & $ 0.76074\,(0.00145) $\\
3 & 3 & $ 0.47449\,(0.00048) $ & $ 0.53493\,(0.00073) $ & $ 0.58625\,(0.00112) $ & $ 0.64421\,(0.00124) $ & $ 0.68102\,(0.00171) $\\
[4pt]
4 & 0 & $ 0.64649 $ & $ 0.71065 $ & $ 0.76150 $ & $ 0.81399 $ & $ 0.84531 $\\
4 & 1 & $ 0.58424\,(0.00053) $ & $ 0.64873\,(0.00059) $ & $ 0.70139\,(0.00076) $ & $ 0.75700\,(0.00112) $ & $ 0.79072\,(0.00127) $\\
4 & 2 & $ 0.52734\,(0.00069) $ & $ 0.59014\,(0.00077) $ & $ 0.64187\,(0.00103) $ & $ 0.69996\,(0.00127) $ & $ 0.73612\,(0.00133) $\\
\bottomrule
\end{tabularx}
\end{table}

\setcounter{table}{4}

\begin{table}[H] \small
\caption{Quantiles of $T_{{\mathcal G}_n}$ with $n_f$ fake $p$-values (continued)}
\vspace{-0.2cm}
		\newcolumntype{C}{>{\centering\arraybackslash}X}
		\begin{tabularx}{\textwidth}{ccXXXXX}
			\toprule
${n}$ & ${n_f}$ &  {0.005} & {0.010} & {0.025} & {0.050} & {0.100} 
\\
\midrule
14 & 0 & $ 0.16183 $ & $ 0.17831 $ & $ 0.20436 $ & $ 0.22848 $ & $ 0.25817 $\\
14 & 1 & $ 0.15528\,(0.00069) $ & $ 0.17118\,(0.00049) $ & $ 0.19608\,(0.00038) $ & $ 0.21908\,(0.00029) $ & $ 0.24773\,(0.00024) $\\
14 & 2 & $ 0.14907\,(0.00061) $ & $ 0.16460\,(0.00048) $ & $ 0.18856\,(0.00031) $ & $ 0.21089\,(0.00026) $ & $ 0.23882\,(0.00024) $\\
14 & 3 & $ 0.14271\,(0.00054) $ & $ 0.15788\,(0.00043) $ & $ 0.18145\,(0.00032) $ & $ 0.20322\,(0.00026) $ & $ 0.23002\,(0.00024) $\\
14 & 4 & $ 0.13722\,(0.00055) $ & $ 0.15175\,(0.00046) $ & $ 0.17483\,(0.00031) $ & $ 0.19583\,(0.00028) $ & $ 0.22127\,(0.00024) $\\
[4pt]
15 & 0 & $ 0.16712 $ & $ 0.18334 $ & $ 0.20888 $ & $ 0.23245 $ & $ 0.26136 $\\
15 & 1 & $ 0.16029\,(0.00070) $ & $ 0.17692\,(0.00049) $ & $ 0.20097\,(0.00042) $ & $ 0.22334\,(0.00028) $ & $ 0.25136\,(0.00023) $\\
\bottomrule
\end{tabularx}
\end{table}

\setcounter{table}{4}

\begin{table}[H] \small
\caption{Quantiles of $T_{{\mathcal G}_n}$ with $n_f$ fake $p$-values (continued)}
 \vspace{-0.2cm}
		\newcolumntype{C}{>{\centering\arraybackslash}X}
		\begin{tabularx}{\textwidth}{ccXXXXX}
			\toprule
${n}$ & ${n_f}$ &  {0.900} & {0.950} & {0.975} & {0.990} & {0.995} 
\\
\midrule
14 & 0 & $ 0.50844 $ & $ 0.54631 $ & $ 0.57885 $ & $ 0.61603 $ & $ 0.64079 $\\
14 & 1 & $ 0.49210\,(0.00041) $ & $ 0.52997\,(0.00052) $ & $ 0.56278\,(0.00052) $ & $ 0.59926\,(0.00057) $ & $ 0.62356\,(0.00076) $\\
14 & 2 & $ 0.47653\,(0.00036) $ & $ 0.51375\,(0.00045) $ & $ 0.54565\,(0.00055) $ & $ 0.58192\,(0.00071) $ & $ 0.60611\,(0.00084) $\\
14 & 3 & $ 0.46124\,(0.00034) $ & $ 0.49738\,(0.00040) $ & $ 0.52834\,(0.00055) $ & $ 0.56577\,(0.00067) $ & $ 0.58953\,(0.00096) $\\
14 & 4 & $ 0.44617\,(0.00031) $ & $ 0.48169\,(0.00041) $ & $ 0.51266\,(0.00045) $ & $ 0.54877\,(0.00073) $ & $ 0.57402\,(0.00090) $\\
[4pt]
15 & 0 & $ 0.50326 $ & $ 0.53987 $ & $ 0.57138 $ & $ 0.60747 $ & $ 0.63156 $\\
15 & 1 & $ 0.48846\,(0.00036) $ & $ 0.52437\,(0.00044) $ & $ 0.55598\,(0.00047) $ & $ 0.59191\,(0.00059) $ & $ 0.61517\,(0.00071) $\\
\bottomrule
\end{tabularx}
\end{table}

\setcounter{table}{4}

\begin{table}[H] \small
\caption{Quantiles of $T_{{\mathcal G}_n}$ with $n_f$ fake $p$-values (continued)}
 \vspace{-0.2cm}
		\newcolumntype{C}{>{\centering\arraybackslash}X}
		\begin{tabularx}{\textwidth}{ccXXXXX}
			\toprule
${n}$ & ${n_f}$ &  {0.005} & {0.010} & {0.025} & {0.050} & {0.100} 
\\
\midrule
21 & 0 & $ 0.19189 $ & $ 0.20673 $ & $ 0.22972 $ & $ 0.25060 $ & $ 0.27586 $\\
21 & 1 & $ 0.18608\,(0.00056) $ & $ 0.20148\,(0.00043) $ & $ 0.22371\,(0.00036) $ & $ 0.24398\,(0.00032) $ & $ 0.26869\,(0.00024) $\\
21 & 2 & $ 0.18265\,(0.00048) $ & $ 0.19631\,(0.00037) $ & $ 0.21835\,(0.00032) $ & $ 0.23810\,(0.00028) $ & $ 0.26199\,(0.00023) $\\
21 & 3 & $ 0.17746\,(0.00048) $ & $ 0.19096\,(0.00039) $ & $ 0.21237\,(0.00032) $ & $ 0.23199\,(0.00028) $ & $ 0.25550\,(0.00024) $\\
21 & 4 & $ 0.17249\,(0.00049) $ & $ 0.18615\,(0.00039) $ & $ 0.20688\,(0.00031) $ & $ 0.22620\,(0.00026) $ & $ 0.24918\,(0.00022) $\\
21 & 5 & $ 0.16822\,(0.00044) $ & $ 0.18143\,(0.00042) $ & $ 0.20181\,(0.00033) $ & $ 0.22061\,(0.00025) $ & $ 0.24294\,(0.00022) $\\
21 & 6 & $ 0.16394\,(0.00050) $ & $ 0.17632\,(0.00038) $ & $ 0.19646\,(0.00032) $ & $ 0.21477\,(0.00024) $ & $ 0.23690\,(0.00022) $\\
\bottomrule
\end{tabularx}
\end{table}

\setcounter{table}{4}

\begin{table}[H] \small
\caption{Quantiles of $T_{{\mathcal G}_n}$ with $n_f$ fake $p$-values (continued)}
 \vspace{-0.2cm}
		\newcolumntype{C}{>{\centering\arraybackslash}X}
		\begin{tabularx}{\textwidth}{ccXXXXX}
			\toprule
${n}$ & ${n_f}$ &  {0.900} & {0.950} & {0.975} & {0.990} & {0.995} 
\\
\midrule
21 & 0 & $ 0.48070 $ & $ 0.51166 $ & $ 0.53847 $ & $ 0.56944 $ & $ 0.59031 $\\
21 & 1 & $ 0.46953\,(0.00034) $ & $ 0.50021\,(0.00041) $ & $ 0.52636\,(0.00042) $ & $ 0.55695\,(0.00067) $ & $ 0.57836\,(0.00085) $\\
21 & 2 & $ 0.45932\,(0.00037) $ & $ 0.48943\,(0.00034) $ & $ 0.51582\,(0.00047) $ & $ 0.54654\,(0.00062) $ & $ 0.56668\,(0.00073) $\\
21 & 3 & $ 0.44937\,(0.00031) $ & $ 0.47916\,(0.00038) $ & $ 0.50498\,(0.00045) $ & $ 0.53545\,(0.00058) $ & $ 0.55692\,(0.00081) $\\
21 & 4 & $ 0.43973\,(0.00028) $ & $ 0.46900\,(0.00034) $ & $ 0.49518\,(0.00039) $ & $ 0.52505\,(0.00052) $ & $ 0.54560\,(0.00077) $\\
21 & 5 & $ 0.43003\,(0.00028) $ & $ 0.45951\,(0.00037) $ & $ 0.48500\,(0.00044) $ & $ 0.51425\,(0.00055) $ & $ 0.53463\,(0.00076) $\\
21 & 6 & $ 0.42099\,(0.00027) $ & $ 0.44991\,(0.00034) $ & $ 0.47471\,(0.00039) $ & $ 0.50387\,(0.00062) $ & $ 0.52457\,(0.00070) $\\
\bottomrule
\end{tabularx}
\end{table}



\vspace*{-0.5cm}
\section*[{Minimum of geometric means statistic  $\boldsymbol {T_{\min\{{\mathcal G_n},{\mathcal G_n^*}\}}(P_1,\dots,P_n)=\min\Big\{\Big(\prod_{k=1}^n P_k\Big)^{1/n}, \Big(\prod_{k=1}^n(1-P_k)\Big)^{1/n}  \Big\}} $}]{Minimum of geometric means statistic\\  $\boldsymbol {T_{\min\{{\mathcal G_n},{\mathcal G_n^*}\}}(P_1,\dots,P_n)=\min\Big\{\Big(\prod_{k=1}^n P_k\Big)^{1/n}, \Big(\prod_{k=1}^n(1-P_k)\Big)^{1/n}  \Big\}} $}

\vspace*{-0.25cm}

\setcounter{table}{5}
\begin{table}[H]\small
\caption{Quantiles of $T_{\min\{{\mathcal G_n},{\mathcal G_n^*}\}}$ with $n_f$ fake $p$-values [estimate\,(standard \mbox{error})]}\label{Table.mGM}
\vspace{-0.25cm}
		\newcolumntype{C}{>{\centering\arraybackslash}X}
\begin{tabularx}{\textwidth}{ccXXXXX}
			\toprule
${n}$ & ${n_f}$ &  {0.005} & {0.010} & {0.025} & {0.050} & {0.100} 
\\
\midrule
3 & 0 & $ 0.03449\,(0.00042) $ & $ 0.04537\,(0.00035) $ & $ 0.06676\,(0.00033) $ & $ 0.08980\,(0.00032) $ & $ 0.12234\,(0.00027) $\\
3 & 1 & $ 0.03390\,(0.00039) $ & $ 0.04530\,(0.00042) $ & $ 0.06666\,(0.00032) $ & $ 0.08962\,(0.00030) $ & $ 0.12201\,(0.00030) $\\
3 & 2 & $ 0.03004\,(0.00040) $ & $ 0.04003\,(0.00030) $ & $ 0.05915\,(0.00030) $ & $ 0.07998\,(0.00031) $ & $ 0.11076\,(0.00033) $\\
3 & 3 & $ 0.04056\,(0.00041) $ & $ 0.05170\,(0.00036) $ & $ 0.07104\,(0.00027) $ & $ 0.09217\,(0.00027) $ & $ 0.12108\,(0.00030) $\\
[4pt]
4 & 0 & $ 0.05127\,(0.00049) $ & $ 0.06470\,(0.00041) $ & $ 0.08771\,(0.00039) $ & $ 0.11188\,(0.00036) $ & $ 0.14402\,(0.00029) $\\
4 & 1 & $ 0.05088\,(0.00042) $ & $ 0.06411\,(0.00045) $ & $ 0.08792\,(0.00034) $ & $ 0.11150\,(0.00037) $ & $ 0.14353\,(0.00031) $\\
4 & 2 & $ 0.04669\,(0.00041) $ & $ 0.05874\,(0.00034) $ & $ 0.08075\,(0.00032) $ & $ 0.10400\,(0.00031) $ & $ 0.13508\,(0.00036) $\\
\bottomrule
\end{tabularx}
\end{table}

\setcounter{table}{5}

\begin{table}[H] \small
\caption{Quantiles of $T_{\min\{{\mathcal G_n},{\mathcal G_n^*}\}}$ with $n_f$ fake $p$-values (continued)}
\vspace{-0.2cm} 
		\newcolumntype{C}{>{\centering\arraybackslash}X}
		\begin{tabularx}{\textwidth}{ccXXXXX}
			\toprule
${n}$ & ${n_f}$ &  {0.900} & {0.950} & {0.975} & {0.990} & {0.995} 
\\
\midrule
3 & 0 & $ 0.41468\,(0.00029) $ & $ 0.44134\,(0.00024) $ & $ 0.45920\,(0.00028) $ & $ 0.47490\,(0.00026) $ & $ 0.48216\,(0.00026) $\\
3 & 1 & $ 0.41442\,(0.00029) $ & $ 0.44146\,(0.00025) $ & $ 0.45939\,(0.00025) $ & $ 0.47483\,(0.00024) $ & $ 0.48234\,(0.00028) $\\
3 & 2 & $ 0.41252\,(0.00024) $ & $ 0.44042\,(0.00028) $ & $ 0.45893\,(0.00023) $ & $ 0.47455\,(0.00022) $ & $ 0.48189\,(0.00022) $\\
3 & 3 & $ 0.40166\,(0.00028) $ & $ 0.42929\,(0.00032) $ & $ 0.44839\,(0.00031) $ & $ 0.46553\,(0.00029) $ & $ 0.47430\,(0.00034) $\\
[4pt]
4 & 0 & $ 0.40681\,(0.00019) $ & $ 0.43223\,(0.00024) $ & $ 0.44978\,(0.00024) $ & $ 0.46612\,(0.00025) $ & $ 0.47452\,(0.00027) $\\
4 & 1 & $ 0.40679\,(0.00025) $ & $ 0.43220\,(0.00029) $ & $ 0.45029\,(0.00027) $ & $ 0.46665\,(0.00028) $ & $ 0.47537\,(0.00028) $\\
4 & 2 & $ 0.40517\,(0.00023) $ & $ 0.43142\,(0.00024) $ & $ 0.44989\,(0.00023) $ & $ 0.46628\,(0.00026) $ & $ 0.47472\,(0.00031) $\\
\bottomrule
\end{tabularx}
\end{table}

\setcounter{table}{5}

\begin{table}[H] \small
\caption{Quantiles of $T_{\min\{{\mathcal G_n},{\mathcal G_n^*}\}}$ with $n_f$ fake $p$-values (continued)}
\vspace{-0.2cm}
		\newcolumntype{C}{>{\centering\arraybackslash}X}
		\begin{tabularx}{\textwidth}{ccXXXXX}
			\toprule
${n}$ & ${n_f}$ &  {0.005} & {0.010} & {0.025} & {0.050} & {0.100} 
\\
\midrule
14 & 0 & $ 0.14730\,(0.00055) $ & $ 0.16162\,(0.00044) $ & $ 0.18418\,(0.00038) $ & $ 0.20425\,(0.00030) $ & $ 0.22834\,(0.00025) $\\
14 & 1 & $ 0.14747\,(0.00068) $ & $ 0.16181\,(0.00049) $ & $ 0.18402\,(0.00033) $ & $ 0.20413\,(0.00027) $ & $ 0.22817\,(0.00023) $\\
14 & 2 & $ 0.14512\,(0.00060) $ & $ 0.15984\,(0.00045) $ & $ 0.18190\,(0.00031) $ & $ 0.20211\,(0.00025) $ & $ 0.22627\,(0.00019) $\\
14 & 3 & $ 0.14103\,(0.00053) $ & $ 0.15573\,(0.00040) $ & $ 0.17812\,(0.00030) $ & $ 0.19845\,(0.00025) $ & $ 0.22293\,(0.00022) $\\
14 & 4 & $ 0.13667\,(0.00055) $ & $ 0.15086\,(0.00045) $ & $ 0.17335\,(0.00032) $ & $ 0.19366\,(0.00027) $ & $ 0.21772\,(0.00021) $\\
[4pt]
15 & 0 & $ 0.15268\,(0.00054) $ & $ 0.16704\,(0.00046) $ & $ 0.18906\,(0.00039) $ & $ 0.20898\,(0.00027) $ & $ 0.23258\,(0.00021) $\\
15 & 1 & $ 0.15259\,(0.00062) $ & $ 0.16721\,(0.00047) $ & $ 0.18890\,(0.00038) $ & $ 0.20865\,(0.00027) $ & $ 0.23210\,(0.00022) $\\
\bottomrule
\end{tabularx}
\end{table}

\setcounter{table}{5}

\begin{table}[H] \small
\caption{Quantiles of $T_{\min\{{\mathcal G_n},{\mathcal G_n^*}\}}$ with $n_f$ fake $p$-values (continued)}
\vspace{-0.2cm}
		\newcolumntype{C}{>{\centering\arraybackslash}X}
		\begin{tabularx}{\textwidth}{ccXXXXX}
			\toprule
${n}$ & ${n_f}$ &  {0.900} & {0.950} & {0.975} & {0.990} & {0.995} 
\\
\midrule
14 & 0 & $ 0.38504\,(0.00017) $ & $ 0.40153\,(0.00022) $ & $ 0.41435\,(0.00022) $ & $ 0.42824\,(0.00025) $ & $ 0.43693\,(0.00032) $\\
14 & 1 & $ 0.38499\,(0.00015) $ & $ 0.40130\,(0.00018) $ & $ 0.41419\,(0.00018) $ & $ 0.42794\,(0.00020) $ & $ 0.43635\,(0.00030) $\\
14 & 2 & $ 0.38443\,(0.00018) $ & $ 0.40085\,(0.00017) $ & $ 0.41404\,(0.00019) $ & $ 0.42765\,(0.00023) $ & $ 0.43620\,(0.00024) $\\
14 & 3 & $ 0.38368\,(0.00016) $ & $ 0.40060\,(0.00019) $ & $ 0.41373\,(0.00018) $ & $ 0.42767\,(0.00022) $ & $ 0.43592\,(0.00031) $\\
14 & 4 & $ 0.38235\,(0.00017) $ & $ 0.39956\,(0.00018) $ & $ 0.41320\,(0.00023) $ & $ 0.42725\,(0.00026) $ & $ 0.43591\,(0.00032) $\\
[4pt]
15 & 0 & $ 0.38450\,(0.00018) $ & $ 0.40033\,(0.00018) $ & $ 0.41307\,(0.00020) $ & $ 0.42624\,(0.00023) $ & $ 0.43483\,(0.00030) $\\
15 & 1 & $ 0.38436\,(0.00015) $ & $ 0.40018\,(0.00018) $ & $ 0.41299\,(0.00019) $ & $ 0.42676\,(0.00021) $ & $ 0.43502\,(0.00032) $\\
\bottomrule
\end{tabularx}
\end{table}

\setcounter{table}{5}

\begin{table}[H] \small
\caption{Quantiles of $T_{\min\{{\mathcal G_n},{\mathcal G_n^*}\}}$ with $n_f$ fake $p$-values (continued)}
\vspace{-0.2cm} 
		\newcolumntype{C}{>{\centering\arraybackslash}X}
		\begin{tabularx}{\textwidth}{ccXXXXX}
			\toprule
${n}$ & ${n_f}$ &  {0.005} & {0.010} & {0.025} & {0.050} & {0.100} 
\\
\midrule
21 & 0 & $ 0.17928\,(0.00044) $ & $ 0.19251\,(0.00038) $ & $ 0.21228\,(0.00026) $ & $ 0.23017\,(0.00024) $ & $ 0.25081\,(0.00022) $\\
21 & 1 & $ 0.17863\,(0.00049) $ & $ 0.19197\,(0.00039) $ & $ 0.21235\,(0.00031) $ & $ 0.22995\,(0.00023) $ & $ 0.25076\,(0.00022) $\\
21 & 2 & $ 0.17762\,(0.00045) $ & $ 0.19076\,(0.00035) $ & $ 0.21108\,(0.00029) $ & $ 0.22873\,(0.00025) $ & $ 0.24976\,(0.00019) $\\
21 & 3 & $ 0.17509\,(0.00047) $ & $ 0.18793\,(0.00033) $ & $ 0.20824\,(0.00024) $ & $ 0.22621\,(0.00025) $ & $ 0.24741\,(0.00022) $\\
21 & 4 & $ 0.17118\,(0.00048) $ & $ 0.18463\,(0.00035) $ & $ 0.20456\,(0.00028) $ & $ 0.22287\,(0.00024) $ & $ 0.24411\,(0.00021) $\\
21 & 5 & $ 0.16785\,(0.00043) $ & $ 0.18079\,(0.00043) $ & $ 0.20066\,(0.00032) $ & $ 0.21874\,(0.00026) $ & $ 0.23993\,(0.00020) $\\
21 & 6 & $ 0.16368\,(0.00051) $ & $ 0.17605\,(0.00036) $ & $ 0.19589\,(0.00032) $ & $ 0.21392\,(0.00023) $ & $ 0.23526\,(0.00020) $\\
\bottomrule
\end{tabularx}
\end{table}

\setcounter{table}{5}

\begin{table}[H] \small
\caption{Quantiles of $T_{\min\{{\mathcal G_n},{\mathcal G_n^*}\}}$  with $n_f$ fake $p$-values (continued)}
\vspace{-0.2cm} 
		\newcolumntype{C}{>{\centering\arraybackslash}X}
		\begin{tabularx}{\textwidth}{ccXXXXX}
			\toprule
${n}$ & ${n_f}$ &  {0.900} & {0.950} & {0.975} & {0.990} & {0.995} 
\\
\midrule
21 & 0 & $ 0.38088\,(0.00014) $ & $ 0.39486\,(0.00016) $ & $ 0.40618\,(0.00020) $ & $ 0.41797\,(0.00023) $ & $ 0.42525\,(0.00032) $\\
21 & 1 & $ 0.38095\,(0.00014) $ & $ 0.39493\,(0.00014) $ & $ 0.40626\,(0.00017) $ & $ 0.41784\,(0.00021) $ & $ 0.42542\,(0.00028) $\\
21 & 2 & $ 0.38062\,(0.00013) $ & $ 0.39491\,(0.00014) $ & $ 0.40616\,(0.00017) $ & $ 0.41816\,(0.00023) $ & $ 0.42580\,(0.00025) $\\
21 & 3 & $ 0.38007\,(0.00015) $ & $ 0.39447\,(0.00017) $ & $ 0.40572\,(0.00022) $ & $ 0.41779\,(0.00026) $ & $ 0.42555\,(0.00028) $\\
21 & 4 & $ 0.37940\,(0.00013) $ & $ 0.39397\,(0.00014) $ & $ 0.40533\,(0.00019) $ & $ 0.41787\,(0.00026) $ & $ 0.42540\,(0.00029) $\\
21 & 5 & $ 0.37844\,(0.00013) $ & $ 0.39311\,(0.00016) $ & $ 0.40460\,(0.00022) $ & $ 0.41664\,(0.00025) $ & $ 0.42435\,(0.00030) $\\
21 & 6 & $ 0.37708\,(0.00015) $ & $ 0.39203\,(0.00015) $ & $ 0.40406\,(0.00019) $ & $ 0.41603\,(0.00025) $ & $ 0.42425\,(0.00030) $\\
\bottomrule
\end{tabularx}
\end{table}



\section*{Wilson's harmonic mean statistic   $\boldsymbol{T_{{{\mathcal H}_n}}(P_1,\dotsc,P_n) =  {n}/{ \sum_{k=1}^n \frac{1}{P_k}} }$}

\vspace*{-0.5cm}

\setcounter{table}{6}

\begin{table}[H]\small
\caption{Quantiles of $T_{{{\mathcal H}_n}}$ with $n_f$ fake $p$-values [estimate\,(standard error)]}\label{Table.Wilson} 
\vspace{-0.4cm}
		\newcolumntype{C}{>{\centering\arraybackslash}X}
\begin{tabularx}{\textwidth}{ccXXXXX}
			\toprule
${n}$ & ${n_f}$ &  {0.005} & {0.010} & {0.025} & {0.050} & {0.100} 
\\
\midrule
3 & 0 & $ 0.00488\,(0.00016) $ & $ 0.00962\,(0.00022) $ & $ 0.02343\,(0.00030) $ & $ 0.04436\,(0.00037) $ & $ 0.08318\,(0.00046) $\\
3 & 1 & $ 0.00366\,(0.00011) $ & $ 0.00712\,(0.00014) $ & $ 0.01745\,(0.00021) $ & $ 0.03339\,(0.00023) $ & $ 0.06319\,(0.00035) $\\
3 & 2 & $ 0.00297\,(0.00007) $ & $ 0.00571\,(0.00011) $ & $ 0.01394\,(0.00016) $ & $ 0.02686\,(0.00023) $ & $ 0.05071\,(0.00026) $\\
3 & 3 & $ 0.00281\,(0.00007) $ & $ 0.00534\,(0.00010) $ & $ 0.01293\,(0.00014) $ & $ 0.02506\,(0.00018) $ & $ 0.04647\,(0.00027) $\\
[4pt]
4 & 0 & $ 0.00481\,(0.00017) $ & $ 0.00954\,(0.00022) $ & $ 0.02304\,(0.00028) $ & $ 0.04301\,(0.00034) $ & $ 0.07999\,(0.00044) $\\
4 & 1 & $ 0.00399\,(0.00012) $ & $ 0.00769\,(0.00017) $ & $ 0.01849\,(0.00024) $ & $ 0.03495\,(0.00026) $ & $ 0.06503\,(0.00036) $\\
4 & 2 & $ 0.00328\,(0.00009) $ & $ 0.00623\,(0.00013) $ & $ 0.01517\,(0.00019) $ & $ 0.02922\,(0.00024) $ & $ 0.05434\,(0.00026) $\\
\bottomrule
\end{tabularx}
\end{table}

\setcounter{table}{6}

\begin{table}[H] \small
\caption{Quantiles of $T_{{{\mathcal H}_n}}$ with $n_f$ fake $p$-values (continued)}
\vspace{-0.2cm}
		\newcolumntype{C}{>{\centering\arraybackslash}X}
		\begin{tabularx}{\textwidth}{ccXXXXX}
			\toprule
${n}$ & ${n_f}$ &  {0.900} & {0.950} & {0.975} & {0.990} & {0.995} 
\\
\midrule

3 & 0 & $ 0.67225\,(0.00072) $ & $ 0.74918\,(0.00087) $ & $ 0.80705\,(0.00089) $ & $ 0.86247\,(0.00102) $ & $ 0.89202\,(0.00117) $\\
3 & 1 & $ 0.57791\,(0.00093) $ & $ 0.65860\,(0.00091) $ & $ 0.71984\,(0.00091) $ & $ 0.78380\,(0.00103) $ & $ 0.82298\,(0.00144) $\\
3 & 2 & $ 0.49881\,(0.00085) $ & $ 0.57707\,(0.00100) $ & $ 0.64026\,(0.00117) $ & $ 0.70732\,(0.00142) $ & $ 0.75059\,(0.00154) $\\
3 & 3 & $ 0.42882\,(0.00059) $ & $ 0.49757\,(0.00085) $ & $ 0.55640\,(0.00118) $ & $ 0.62198\,(0.00139) $ & $ 0.66185\,(0.00179) $\\
[4pt]
4 & 0 & $ 0.61361\,(0.00066) $ & $ 0.68798\,(0.00082) $ & $ 0.74774\,(0.00087) $ & $ 0.80678\,(0.00095) $ & $ 0.84050\,(0.00123) $\\
4 & 1 & $ 0.54231\,(0.00066) $ & $ 0.61705\,(0.00078) $ & $ 0.67837\,(0.00087) $ & $ 0.74095\,(0.00122) $ & $ 0.77885\,(0.00143) $\\
4 & 2 & $ 0.48158\,(0.00078) $ & $ 0.55387\,(0.00085) $ & $ 0.61429\,(0.00115) $ & $ 0.67987\,(0.00153) $ & $ 0.71988\,(0.00156) $\\
\bottomrule
\end{tabularx}
\end{table}

\setcounter{table}{6}

\begin{table}[H] \small
\caption{Quantiles of $T_{{{\mathcal H}_n}}$ with $n_f$ fake $p$-values (continued)}
\vspace{-0.2cm}
		\newcolumntype{C}{>{\centering\arraybackslash}X}
		\begin{tabularx}{\textwidth}{ccXXXXX}
			\toprule
${n}$ & ${n_f}$ &  {0.005} & {0.010} & {0.025} & {0.050} & {0.100} 
\\
\midrule
14 & 0 & $ 0.00491\,(0.00014) $ & $ 0.00940\,(0.00022) $ & $ 0.02163\,(0.00027) $ & $ 0.03978\,(0.00029) $ & $ 0.06991\,(0.00033) $\\
14 & 1 & $ 0.00455\,(0.00013) $ & $ 0.00871\,(0.00018) $ & $ 0.02020\,(0.00026) $ & $ 0.03714\,(0.00028) $ & $ 0.06562\,(0.00030) $\\
14 & 2 & $ 0.00423\,(0.00010) $ & $ 0.00815\,(0.00016) $ & $ 0.01884\,(0.00023) $ & $ 0.03464\,(0.00022) $ & $ 0.06166\,(0.00028) $\\
14 & 3 & $ 0.00404\,(0.00011) $ & $ 0.00770\,(0.00015) $ & $ 0.01769\,(0.00021) $ & $ 0.03268\,(0.00025) $ & $ 0.05809\,(0.00028) $\\
14 & 4 & $ 0.00382\,(0.00010) $ & $ 0.00730\,(0.00013) $ & $ 0.01676\,(0.00020) $ & $ 0.03103\,(0.00026) $ & $ 0.05512\,(0.00025) $\\
[4pt]
15 & 0 & $ 0.00499\,(0.00014) $ & $ 0.00942\,(0.00024) $ & $ 0.02173\,(0.00029) $ & $ 0.03959\,(0.00029) $ & $ 0.06950\,(0.00033) $\\
15 & 1 & $ 0.00464\,(0.00012) $ & $ 0.00880\,(0.00019) $ & $ 0.02030\,(0.00025) $ & $ 0.03696\,(0.00026) $ & $ 0.06538\,(0.00030) $\\
\bottomrule
\end{tabularx}
\end{table}

\setcounter{table}{6}

\begin{table}[H] \small
\caption{Quantiles of $T_{{{\mathcal H}_n}}$ with $n_f$ fake $p$-values (continued)}
 \vspace{-0.2cm}
		\newcolumntype{C}{>{\centering\arraybackslash}X}
		\begin{tabularx}{\textwidth}{ccXXXXX}
			\toprule
${n}$ & ${n_f}$ &  {0.900} & {0.950} & {0.975} & {0.990} & {0.995} 
\\
\midrule
14 & 0 & $ 0.41340\,(0.00043) $ & $ 0.46483\,(0.00056) $ & $ 0.50865\,(0.00066) $ & $ 0.55723\,(0.00092) $ & $ 0.58869\,(0.00115) $\\
14 & 1 & $ 0.39479\,(0.00050) $ & $ 0.44572\,(0.00050) $ & $ 0.48848\,(0.00071) $ & $ 0.53678\,(0.00085) $ & $ 0.56839\,(0.00089) $\\
14 & 2 & $ 0.37810\,(0.00044) $ & $ 0.42715\,(0.00059) $ & $ 0.46896\,(0.00060) $ & $ 0.51677\,(0.00092) $ & $ 0.54891\,(0.00106) $\\
14 & 3 & $ 0.36159\,(0.00039) $ & $ 0.40959\,(0.00053) $ & $ 0.45087\,(0.00063) $ & $ 0.49813\,(0.00095) $ & $ 0.52992\,(0.00128) $\\
14 & 4 & $ 0.34631\,(0.00044) $ & $ 0.39299\,(0.00049) $ & $ 0.43329\,(0.00051) $ & $ 0.48142\,(0.00091) $ & $ 0.51229\,(0.00124) $\\
[4pt]
15 & 0 & $ 0.40439\,(0.00043) $ & $ 0.45444\,(0.00060) $ & $ 0.49710\,(0.00079) $ & $ 0.54465\,(0.00099) $ & $ 0.57605\,(0.00121) $\\
15 & 1 & $ 0.38814\,(0.00040) $ & $ 0.43706\,(0.00050) $ & $ 0.47908\,(0.00063) $ & $ 0.52706\,(0.00074) $ & $ 0.55728\,(0.00098) $\\
\bottomrule
\end{tabularx}
\end{table}

\setcounter{table}{6}

\begin{table}[H] \small
\caption{Quantiles of $T_{{{\mathcal H}_n}}$ with $n_f$ fake $p$-values (continued)}
\vspace{-0.2cm}
		\newcolumntype{C}{>{\centering\arraybackslash}X}
		\begin{tabularx}{\textwidth}{ccXXXXX}
			\toprule
${n}$ & ${n_f}$ &  {0.005} & {0.010} & {0.025} & {0.050} & {0.100} 
\\
\midrule
21 & 0 & $ 0.00487\,(0.00012) $ & $ 0.00934\,(0.00018) $ & $ 0.02158\,(0.00026) $ & $ 0.03893\,(0.00030) $ & $ 0.06746\,(0.00031) $\\
21 & 1 & $ 0.00470\,(0.00013) $ & $ 0.00896\,(0.00015) $ & $ 0.02065\,(0.00023) $ & $ 0.03717\,(0.00027) $ & $ 0.06440\,(0.00030) $\\
21 & 2 & $ 0.00456\,(0.00012) $ & $ 0.00848\,(0.00012) $ & $ 0.01973\,(0.00022) $ & $ 0.03553\,(0.00024) $ & $ 0.06167\,(0.00027) $\\
21 & 3 & $ 0.00435\,(0.00012) $ & $ 0.00818\,(0.00015) $ & $ 0.01883\,(0.00020) $ & $ 0.03402\,(0.00025) $ & $ 0.05917\,(0.00028) $\\
21 & 4 & $ 0.00423\,(0.00012) $ & $ 0.00789\,(0.00014) $ & $ 0.01829\,(0.00021) $ & $ 0.03292\,(0.00025) $ & $ 0.05696\,(0.00028) $\\
21 & 5 & $ 0.00407\,(0.00011) $ & $ 0.00767\,(0.00014) $ & $ 0.01760\,(0.00021) $ & $ 0.03163\,(0.00026) $ & $ 0.05498\,(0.00029) $\\
21 & 6 & $ 0.00392\,(0.00010) $ & $ 0.00741\,(0.00015) $ & $ 0.01697\,(0.00021) $ & $ 0.03041\,(0.00024) $ & $ 0.05314\,(0.00029) $\\
\bottomrule
\end{tabularx}
\end{table}

\setcounter{table}{6}

\begin{table}[H]\small
\caption{Quantiles of $T_{{{\mathcal H}_n}}$ with $n_f$ fake $p$-values (continued)}
 \vspace{-0.2cm}
		\newcolumntype{C}{>{\centering\arraybackslash}X}
		\begin{tabularx}{\textwidth}{ccXXXXX}
			\toprule
${n}$ & ${n_f}$ &  {0.900} & {0.950} & {0.975} & {0.990} & {0.995} 
\\
\midrule
21 & 0 & $ 0.36573\,(0.00042) $ & $ 0.41000\,(0.00053) $ & $ 0.44784\,(0.00069) $ & $ 0.48939\,(0.00096) $ & $ 0.51801\,(0.00125) $\\
21 & 1 & $ 0.35444\,(0.00045) $ & $ 0.39765\,(0.00052) $ & $ 0.43473\,(0.00066) $ & $ 0.47649\,(0.00088) $ & $ 0.50407\,(0.00098) $\\
21 & 2 & $ 0.34372\,(0.00042) $ & $ 0.38602\,(0.00047) $ & $ 0.42272\,(0.00059) $ & $ 0.46394\,(0.00076) $ & $ 0.49193\,(0.00106) $\\
21 & 3 & $ 0.33338\,(0.00046) $ & $ 0.37502\,(0.00048) $ & $ 0.41034\,(0.00062) $ & $ 0.45181\,(0.00090) $ & $ 0.48023\,(0.00105) $\\
21 & 4 & $ 0.32380\,(0.00035) $ & $ 0.36452\,(0.00044) $ & $ 0.39907\,(0.00055) $ & $ 0.44012\,(0.00075) $ & $ 0.46693\,(0.00108) $\\
21 & 5 & $ 0.31406\,(0.00036) $ & $ 0.35419\,(0.00042) $ & $ 0.38875\,(0.00049) $ & $ 0.42814\,(0.00071) $ & $ 0.45459\,(0.00098) $\\
21 & 6 & $ 0.30517\,(0.00036) $ & $ 0.34461\,(0.00043) $ & $ 0.37843\,(0.00056) $ & $ 0.41641\,(0.00066) $ & $ 0.44384\,(0.00111) $\\
\bottomrule
\end{tabularx}
\end{table}



\vspace*{-0.5cm}

\section*{Edgington's  arithmetic mean statistic  $\boldsymbol{T_E(P_1,\dotsc,P_n) =   \frac{1}{n} \sum_{k=1}^n P_k }$}

\vspace*{-0.5cm}

\setcounter{table}{7}

\begin{table}[H]\small
\caption{Quantiles of $T_E$ with $n_f$ fake $p$-values [estimate\,(standard error)]}\label{Table.Edgington}
\vspace{-0.4cm}
		\newcolumntype{C}{>{\centering\arraybackslash}X}
\begin{tabularx}{\textwidth}{ccXXXXX}
			\toprule
${n}$ & ${n_f}$ &  {0.005} & {0.010} & {0.025} & {0.050} & {0.100} 
\\
\midrule
3 & 0 & $ 0.10372\,(0.00088) $ & $ 0.13039\,(0.00080) $ & $ 0.17634\,(0.00069) $ & $ 0.22224\,(0.00059) $ & $ 0.28043\,(0.00051) $\\
3 & 1 & $ 0.08339\,(0.00099) $ & $ 0.10666\,(0.00087) $ & $ 0.14530\,(0.00076) $ & $ 0.18515\,(0.00059) $ & $ 0.23695\,(0.00055) $\\
3 & 2 & $ 0.06763\,(0.00061) $ & $ 0.08604\,(0.00059) $ & $ 0.11874\,(0.00049) $ & $ 0.15231\,(0.00049) $ & $ 0.19714\,(0.00045) $\\
3 & 3 & $ 0.09295\,(0.00062) $ & $ 0.11219\,(0.00054) $ & $ 0.14383\,(0.00046) $ & $ 0.17505\,(0.00047) $ & $ 0.21410\,(0.00038) $\\
[4pt]
4 & 0 & $ 0.14726\,(0.00095) $ & $ 0.17460\,(0.00078) $ & $ 0.21924\,(0.00057) $ & $ 0.26122\,(0.00046) $ & $ 0.31061\,(0.00043) $\\
4 & 1 & $ 0.12734\,(0.00096) $ & $ 0.15115\,(0.00077) $ & $ 0.19193\,(0.00050) $ & $ 0.23124\,(0.00049) $ & $ 0.27767\,(0.00045) $\\
4 & 2 & $ 0.10846\,(0.00064) $ & $ 0.13031\,(0.00068) $ & $ 0.16722\,(0.00061) $ & $ 0.20187\,(0.00052) $ & $ 0.24597\,(0.00041) $\\
\bottomrule
\end{tabularx}
\end{table}

\setcounter{table}{7}

\begin{table}[H] \small
\caption{Quantiles of $T_E$ with $n_f$ fake $p$-values (continued)}
\vspace{-0.2cm} 
		\newcolumntype{C}{>{\centering\arraybackslash}X}
		\begin{tabularx}{\textwidth}{ccXXXXX}
			\toprule
${n}$ & ${n_f}$ &  {0.900} & {0.950} & {0.975} & {0.990} & {0.995} 
\\
\midrule

3 & 0 & $ 0.71897\,(0.00048) $ & $ 0.77728\,(0.00064) $ & $ 0.82462\,(0.00071) $ & $ 0.87166\,(0.00080) $ & $ 0.89761\,(0.00100) $\\
3 & 1 & $ 0.65198\,(0.00059) $ & $ 0.70774\,(0.00064) $ & $ 0.75395\,(0.00076) $ & $ 0.80483\,(0.00093) $ & $ 0.83693\,(0.00122) $\\
3 & 2 & $ 0.58438\,(0.00063) $ & $ 0.63872\,(0.00073) $ & $ 0.68583\,(0.00085) $ & $ 0.73788\,(0.00113) $ & $ 0.77339\,(0.00121) $\\
3 & 3 & $ 0.53948\,(0.00046) $ & $ 0.58740\,(0.00062) $ & $ 0.62833\,(0.00083) $ & $ 0.67439\,(0.00101) $ & $ 0.70562\,(0.00132) $\\
[4pt]
4 & 0 & $ 0.68819\,(0.00044) $ & $ 0.73857\,(0.00054) $ & $ 0.78056\,(0.00061) $ & $ 0.82680\,(0.00070) $ & $ 0.85444\,(0.00108) $\\
4 & 1 & $ 0.63951\,(0.00046) $ & $ 0.68891\,(0.00041) $ & $ 0.72987\,(0.00060) $ & $ 0.77504\,(0.00088) $ & $ 0.80570\,(0.00101) $\\
4 & 2 & $ 0.59030\,(0.00050) $ & $ 0.63879\,(0.00063) $ & $ 0.67952\,(0.00077) $ & $ 0.72579\,(0.00111) $ & $ 0.75666\,(0.00118) $\\
\bottomrule
\end{tabularx}
\end{table}

\setcounter{table}{7}

\begin{table}[H] \small
\caption{Quantiles of $T_E$ with $n_f$ fake $p$-values (continued)}
\vspace{-0.2cm} 
		\newcolumntype{C}{>{\centering\arraybackslash}X}
		\begin{tabularx}{\textwidth}{ccXXXXX}
			\toprule
${n}$ & ${n_f}$ &  {0.005} & {0.010} & {0.025} & {0.050} & {0.100} 
\\
\midrule
14 & 0 & $ 0.30339\,(0.00061) $ & $ 0.32144\,(0.00049) $ & $ 0.34859\,(0.00041) $ & $ 0.37245\,(0.00033) $ & $ 0.39998\,(0.00024) $\\
14 & 1 & $ 0.29416\,(0.00065) $ & $ 0.31204\,(0.00045) $ & $ 0.33900\,(0.00044) $ & $ 0.36219\,(0.00030) $ & $ 0.38936\,(0.00022) $\\
14 & 2 & $ 0.28538\,(0.00067) $ & $ 0.30334\,(0.00053) $ & $ 0.32937\,(0.00033) $ & $ 0.35223\,(0.00028) $ & $ 0.37888\,(0.00024) $\\
14 & 3 & $ 0.27673\,(0.00069) $ & $ 0.29422\,(0.00048) $ & $ 0.31940\,(0.00037) $ & $ 0.34235\,(0.00029) $ & $ 0.36810\,(0.00027) $\\
14 & 4 & $ 0.26833\,(0.00069) $ & $ 0.28487\,(0.00048) $ & $ 0.31052\,(0.00039) $ & $ 0.33243\,(0.00032) $ & $ 0.35794\,(0.00026) $\\
[4pt]
15 & 0 & $ 0.30879\,(0.00068) $ & $ 0.32754\,(0.00045) $ & $ 0.35403\,(0.00036) $ & $ 0.37650\,(0.00032) $ & $ 0.40355\,(0.00024) $\\
15 & 1 & $ 0.30073\,(0.00058) $ & $ 0.31848\,(0.00050) $ & $ 0.34457\,(0.00036) $ & $ 0.36702\,(0.00032) $ & $ 0.39356\,(0.00023) $\\
\bottomrule
\end{tabularx}
\end{table}

\setcounter{table}{7}

\begin{table}[H]\small
\caption{Quantiles of $T_E$ with $n_f$ fake $p$-values (continued)}
\vspace{-0.2cm}
		\newcolumntype{C}{>{\centering\arraybackslash}X}
		%
\end{table}

\setcounter{table}{7}

\begin{table}[H]\small
\caption{Quantiles of $T_E$ with $n_f$ fake $p$-values (continued)}
 \vspace{-0.2cm}
		\newcolumntype{C}{>{\centering\arraybackslash}X}
		%
\end{table}

\setcounter{table}{7}

\begin{table}[H]\small
\caption{Quantiles of $T_E$ with $n_f$ fake $p$-values (continued)}
\vspace{-0.2cm}
		\newcolumntype{C}{>{\centering\arraybackslash}X}
		%
\end{table}



\section*{Tippett's minimum statistic $\boldsymbol{T_T(P_1,\dotsc,P_n)= P_{1:n} }$}

\vspace*{-0.5cm}

\setcounter{table}{8}

\begin{table}[H] \small
\caption{Quantiles of $T_T$ with $n_f$ fake $p$-values}\label{Table.Tippett}
\vspace{-0.4cm}
		\newcolumntype{C}{>{\centering\arraybackslash}X}
		%
\end{table}

\setcounter{table}{8}

\begin{table}[H] \small
\caption{Quantiles of $T_T$ with $n_f$ fake $p$-values (continued)}
 \vspace{-0.2cm}
		\newcolumntype{C}{>{\centering\arraybackslash}X}
		%
\end{table}

\setcounter{table}{8}

\begin{table}[H] \small
\caption{Quantiles of $T_T$ with $n_f$ fake $p$-values (continued)}
\vspace{-0.2cm}
		\newcolumntype{C}{>{\centering\arraybackslash}X}
		%
\end{table}

\setcounter{table}{8}

\begin{table}[H]\small
\caption{Quantiles of $T_T$ with $n_f$ fake $p$-values (continued)}
 \vspace{-0.2cm}
		\newcolumntype{C}{>{\centering\arraybackslash}X}
		%
\end{table}

\setcounter{table}{8}

\begin{table}[H]\small
\caption{Quantiles of $T_T$ with $n_f$ fake $p$-values (continued)}
 \vspace{-0.2cm}
		\newcolumntype{C}{>{\centering\arraybackslash}X}
		%
\end{table}

\setcounter{table}{8}

\begin{table}[H]\small
\caption{Quantiles of $T_T$ with $n_f$ fake $p$-values (continued)}
\vspace{-0.2cm}
		\newcolumntype{C}{>{\centering\arraybackslash}X}
		%
\end{table}


\section*{Wilkinson's maximum statistic $\boldsymbol{T_{W}(P_1,\dotsc,P_n) = P_{n:n}}$}

\vspace*{-0.5cm}

\setcounter{table}{9}

\begin{table}[H]\small
\caption{Quantiles of $T_{W}$ with $n_f$ fake $p$-values [estimate\,(standard error)]}\label{Table.Wilkinson}
\vspace{-0.4cm}
		\newcolumntype{C}{>{\centering\arraybackslash}X}
		%
\end{table}

\setcounter{table}{9}

\begin{table}[H]\small
\caption{Quantiles of $T_{W}$ with $n_f$ fake $p$-values (continued)}
\vspace{-0.2cm}
		\newcolumntype{C}{>{\centering\arraybackslash}X}
		%
\end{table}

\setcounter{table}{9}

\begin{table}[H]\small
\caption{Quantiles of $T_{W}$ with $n_f$ fake $p$-values (continued)}
\vspace{-0.2cm}
		\newcolumntype{C}{>{\centering\arraybackslash}X}
		%
\end{table}

\setcounter{table}{9}

\begin{table}[H]\small
\caption{Quantiles of $T_{W}$ with $n_f$ fake $p$-values (continued)}
\vspace{-0.2cm} 
		\newcolumntype{C}{>{\centering\arraybackslash}X}
		%
\end{table}

\setcounter{table}{9}

\begin{table}[H]\small
\caption{Quantiles of $T_{W}$ with $n_f$ fake $p$-values (continued)}
\vspace{-0.2cm} 
		\newcolumntype{C}{>{\centering\arraybackslash}X}
		%
\end{table}

\setcounter{table}{9}

\begin{table}[H]\small
\caption{Quantiles of $T_{W}$ with $n_f$ fake $p$-values (continued)}
\vspace{-0.2cm} 
		\newcolumntype{C}{>{\centering\arraybackslash}X}
		%
\end{table}
 
\bibliographystyle{alea3}
\bibliography{RefsMetaAnalysis}

\end{sloppy}

\end{document}